\documentclass[structabstract]{aa}  
\usepackage{graphicx}
\usepackage{float} 
\usepackage{morefloats} 
\usepackage{amssymb} 
\usepackage{natbib}
\usepackage{subfigure}
\usepackage{multirow}
\usepackage{threeparttable}
\usepackage[table]{xcolor}

\newcommand{\xte}{\hbox{XTE\,J1701$-$462}}
\newcommand{\mxb}{\hbox{MXB\,1659$-$29}}
\newcommand{\ks}{\hbox{KS\,1731$-$260}}
\newcommand{\exo}{\hbox{EXO\,0748$-$676}}
\newcommand{\exoo}{\hbox{EXO\,1745$-$248}}
\newcommand{\ig}{\hbox{IGR\,J17480$-$2446}}

\newcommand{\q}{$Q_{\rm imp}$}
\newcommand{\tc}{$T_{\rm c}$}
\newcommand{\tb}{$T_{\rm b}$}
\newcommand{\dm}{$\dot{M}$}

\newcommand{\xmm}{XMM-Newton}
\newcommand{\ch}{Chandra}
\newcommand{\sw}{Swift}
\newcommand{\tbts}{$T_{\rm b}$--$T_{\rm s}$}

\newcommand{\tableTop}{\rule{0pt}{2.6ex}}       
\newcommand{\tableBot}{\rule[-1.2ex]{0pt}{0pt}} 

\begin{document}

   \title{Quiescent thermal emission from neutron stars in LMXBs}

   \author{A. Turlione
          \inst{1}, D.N. Aguilera\inst{1,2}
          \and
          J. A. Pons \inst{3}
          }
         \institute{Laboratorio Tandar, CNEA-CONICET, Av. Gral Paz 1499, 1430 San Mart\'in, Buenos Aires, 
 Argentina\\
\and
Deutsches Zentrum f\"ur Luft-und Raumfahrt, DLR-RY, Robert-Hooke Str. 7, 28359 Bremen, Germany\\
 \and 
Departament de F\'{\i}sica Aplicada, Universitat d'Alacant,  Ap. Correus 99, 03080 Alacant, Spain
             }

\date{}
\abstract
{We monitored the quiescent thermal emission from neutron stars in low-mass X-ray
binaries after active periods of intense activity in x-rays (outbursts).}
{The theoretical modeling of the thermal relaxation 
of the neutron star crust may be used to establish constraints on the crust  composition 
and transport properties, depending on the astrophysical scenarios assumed.}
{ We numerically simulated the thermal evolution of the neutron star crust 
and compared them with inferred surface temperatures for five sources: \mxb, \ks, \xte, \exo\, and \ig. 
}
{     We find that the evolution of \mxb, \ks\ and \exo\ can be well 
      described within a deep crustal cooling scenario.
   Conversely, we find that the other two sources can only be explained with models beyond 
   crustal cooling. For the peculiar emission of \xte\ 
   we propose alternative scenarios such as residual accretion during quiescence, 
   additional heat sources in the outer crust,  and/or thermal isolation of 
   the inner crust due to a buried magnetic field. 
   We also explain the very recent reported temperature of \ig\ with
   an additional heat deposition in the outer crust from shallow sources.} 
{}

\titlerunning{Quiescent thermal emission from NSs}
\authorrunning{Turlione et al.}
\maketitle
%
\section{Introduction}
 Neutron star low-mass X-ray binaries (LMXBs) are systems formed by a neutron star (NS)  that accretes matter 
from a low-mass companion star. These systems are most of the time
in a quiescent state where little accretion occurs with an X-ray luminosity $< 10^{34}$\,erg\,s$^{-1}$. 
Periodically, the compact object undergoes an accretion episode with a corresponding increase in 
luminosity of $\sim10^{36}$--$10^{39}$\,erg\,s$^{-1}$. 
The accreted hydrogen- and helium-rich material at rates $\sim10^{15}$--$10^{18}$\,g\,s$^{-1}$
undergoes thermonuclear fusion within hours to days of reaching
the NS surface, releasing $\sim5$~MeV per accreted nucleon.
(see, e.g., \cite{Bildsten1997,Schatz1999} for a seminal work).
The nuclear burning is thermally 
unstable on weakly magnetized NSs (B $\ll 10^{11}$~G) accreting at $\dot{M} < 10^{18}$\,g\,s$^{-1}$ and 
produces energetic ($\sim10^{39}$\,erg) type I X-ray bursts when $\dot{M} < 10^{17}$\,g\,s$^{-1}$.

At the end of an active period, the emission shows a decreasing X-ray activity (quiescent phase) until a new accretion cycle begins.  
Most of the sources accrete for days or weeks, but there are only few of them that show 
unusually long active phases that last for years or decades. 
Recently,  five so-called quasi-persistent sources 
have been monitored for about $10^3$\,days 
after the end of the outburst: 
\mxb\ (\cite{Wijnands2003,Cackett2008}), 
\ks\ (\cite{Wijnands2001,Cackett2010}),  
\exo\ (\cite{Wolff2008,Degenaar2011,DiazTrigo2011, Degenaar2014nja}),  
\xte\ (\cite{Fridriksson2010,Fridriksson2011}), 
and
 \ig\ (\cite{Degenaar2011a}). 
All these sources have been accreting at rates $\simeq 0.01$--$1$ times the Eddington mass accretion rate, 
$\dot M \simeq 10^{18}$\,g\,s$^{-1}$(\cite{Galloway2008,Degenaar2011}). 
The thermal component of the spectra is consistent with an overall decrease in the surface 
temperature of the NS; only for one source, \mxb,  last measurements indicate 
that the star has reached an equilibrium 
temperature\footnote{While this work was being written, a new observation of \mxb\ was reported, the 
temperature of which has not been clearly determined yet 
(see discussion in \cite{Cackett2013});  we did not include it in our analysis.}, but 
for the others there is evidence for continued cooling (\cite{Fridriksson2011, Degenaar2014nja}). 

Theoretical explanations of the origin of the quiescent X-ray emission 
point to the thermal relaxation of the crust. 
Before the active phase it is assumed that the NS is old enough to have an isothermal interior and its surface 
temperature reflects the core temperature. During outbursts, the crust is 
heated up beyond thermal equilibrium by the accretion of matter that compresses the crust and triggers
nuclear reactions.  Once accretion falls to quiescent levels, 
it cools down by thermal radiation  from the surface
(mainly in the X-ray band),   by heat conduction toward the core and consequent neutrino emission as the outer layers return to equilibrium with the interior;  
see the pioneering work by \cite{Brown1998} and \cite{Colpi2001}. 
In quasi-persistent sources  the crust is the region that is heated up because the outburst duration 
($\sim$~yr) is about as long as the crustal 
diffusion timescale.  Nevertheless, observations of one source show that shorter accretion periods of a few months are 
thought to be responsible for the heating of the crust  (see \cite{Degenaar2011b} and analysis on \exo\ below).  
 Typically,  sources accrete for much shorter time, and the heat is generated 
mostly by thermonuclear reactions in the envelope that rapidly diffuses ($\sim$~s,min) and does not  affect
the interior thermal state. 

As a result of this long-term accretion phase, 
the cooling is modified not only by the energy released in the envelope 
(at densities $10^4$--$10^7$g\,cm$^{-3}$) by thermonuclear reactions, 
but also by the energy generated in the inner crust (at $10^{11}$--$10^{13}$g\,cm$^{-3}$) 
by electron captures, neutron emission, 
and density-driven nuclear fusion reactions (pycnonuclear reactions). 
Then, the so-called  deep crustal heating controls the NS cooling in the quiescence phase. 
The rates of these processes have huge 
uncertainties: which particular reaction is taking place and at which density is still unknown. 
Fortunately, this uncertainty does not significantly affect the value of the total heat released 
$Q_{\rm tot}\simeq 1.9$~MeV (\cite{HaenselZdunik2008} hereafter HZ08, \cite{Gupta2007}),
although the spatial distribution of heat sources in the inner crust is uncertain.

In the past decade, the comparison of observational data with cooling models including deep crustal heating allowed 
investigating crust properties and ultra-dense matter processes that
influence the cooling curves \citep{Rutledge2002}. 
Simulations of the crust relaxation after outbursts for \ks\ and \mxb\
(\cite{Shternin2007} and \cite{BrownCumming2009}; hereafter Sht07 and BC09) suggested a
rather high thermal conductivity in the outer crust (which requires a low impurity content), 
in agreement with recent molecular dynamics calculations \citep{Horowitz2009} but in 
contrast with the inefficient crust conductivity necessary for carbon ignition in superbursts 
\citep{Cooper2005,Cumming2006}.  
A recent interpretation confirms the role of the crustal cooling as responsible for the quiescent emission for \mxb, \ks\, and \xte\ and highlights the importance of the outburst duration in the subsequent 
cooling evolution \citep{PageReddy13}.

Many other open questions as well as new observational data  challenge 
these models in several fronts. 
First some of the sources might still be cooling, as indicated by the last observation of \ks\ 
\citep{Fridriksson2011} and  the high temperatures exhibit in \xte. If these sources indeed continue to 
cool, models should account for longer relaxation times for the crust.  
This opens the possibility of revisiting the analysis 
of BC09, who assumed that the 
quiescent emission of the crust levels off with the core.  
Second, some sources might show variability in the thermal component: \xte\ has shown relatively short 
periods of increased temperatures during  an overall cooling evolution. The origin of this variation 
is not clear, and one possible explanation is low-level accretion onto the NS surface during quiescence 
due to the correlated variability observed in the power-law spectral component \citep{Fridriksson2011}. 
Third,  the candidate for crustal cooling  recently detected in the globurar cluster Terzan 5,
\ig, exhibits a higher temperature than the quiescent base level in 2009  \citep{Degenaar2011}. 
More recently, five new observations have been reported \citep{Degenaar2013},  
making its overall cooling even more puzzling.
BC09-type cooling models can only account for these inferred temperatures 
if there is an additional heat generation in the outer crust \citep{Degenaar2011b} whose location and origin is 
unknown\footnote{Another NSs went into quiescence in Terzan 5, \exoo\ \citep{Degenaar2012}, 
but it cannot be  considered for crustal cooling since it lacks 
thermal emission; nevertheless it sets  strong constraints on the properties of the NS core, which 
has efficiently cooled off.}.  
 Other recent theoretical speculations in the outer crust that may affect the crustal cooling  (for instance, \exo) include  heat convection due to 
the chemical separation of light and heavy nuclei  \citep{MedinCumming2014} or a shell with rapid neutrino cooling, which might have more dramatic consequences \citep{Schatz2014}.

In this paper, we aim at revisiting the problem by performing 
time-dependent simulations of the thermal evolution of the NS crust
with deep crustal heating. The main purpose is to use our models to constrain the general properties 
of the NS crust  (e.g.  
the crust thermal conductivity or impurities) by comparing our results with observational data of all
available sources. We also discuss alternative scenarios for the sources that cannot be completely explained
 only by means of deep crustal heating.  

The paper is organized as follows: in Sect.~\ref{sources} we describe 
the five sources \ks, \mxb, \exo, \xte, and 
\ig  and briefly compare their remarkable characteristics.  
In Sect.~\ref{model} we describe the microphysics of the underlying neutron star model 
 and the details for the numerical code.  
In Sect.~\ref{reviewMXB}-\ref{beyond} we test  our cooling simulations in detail
 for the five sources. 
We summarize  in Sect.~\ref{conclusions}.


\section {Sources}
\label{sources}

\begin{figure*}[!ht]
\centering
\subfigure[\textbf{\mxb}, data and fits from \cite{Cackett2008} and \cite{Cackett2013} (open symbols).]
{\includegraphics[scale=0.25,angle=-90]{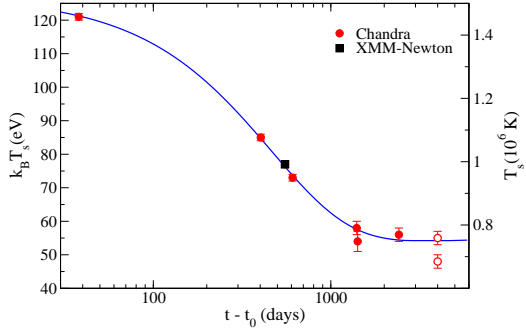}}
\hspace{1cm}
\subfigure[\textbf{\ks}, data and fits from \cite{Cackett2010}.]  
{\includegraphics[scale=0.25,angle=-90]{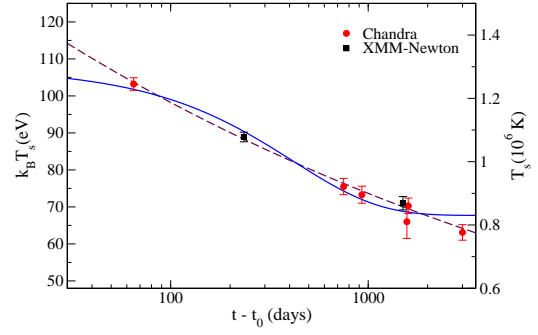}}
\subfigure[\textbf{\exo}, data and fits from \cite{DiazTrigo2011} and \cite{Degenaar2011}.] 
{\includegraphics[scale=0.25,angle=-90]{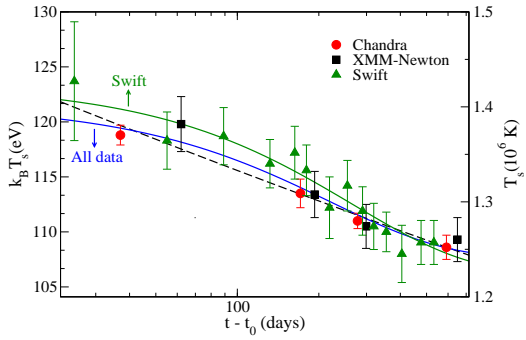}}
\hspace{1cm}
\subfigure[\textbf{\xte}, data and fits from  \cite{Fridriksson2011}. Data with open symbols (XMM-3 and CXO-4) 
were not considered in fits.]
{\includegraphics[scale=0.25,angle=-90]{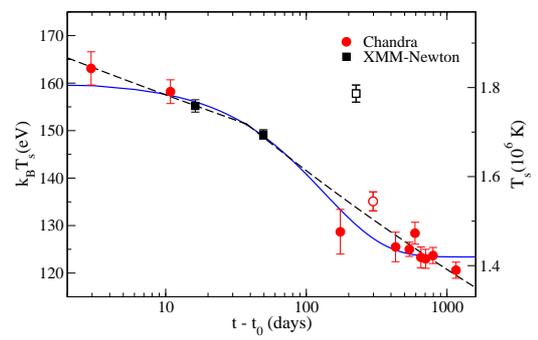}}
\subfigure[\textbf{\ig}, data and fits from \cite{Degenaar2013}.]
{\includegraphics[scale=0.25,angle=-90]{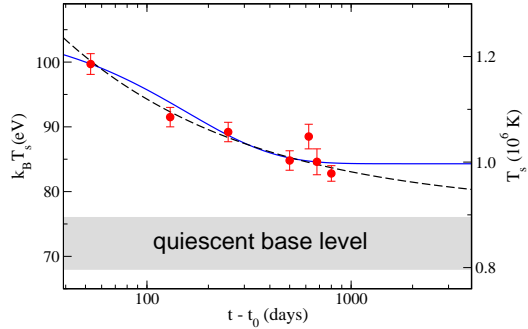}}
\caption{Observational data and corresponding fits taken from the literature. Data from \ch\ (circles), \xmm\ (squares) and \sw\ (triangles)
for all the sources. Exponential decay 
$k_{\rm B} T_s = a\ e^{-(t-t_0)/\tau}+b$ (solid lines),   and (broken) power-laws 
$k_{\rm B}T_s=\alpha(t-t_0)^\beta$ (dashed lines) fits.}
\label{intro}
\end{figure*}

The main observational facts of the five NSs in LMXBs 
detected in quiescence presented below are summarized in Fig.~\ref{intro} and Table~\ref{table:01}. 
\begin{table*}[!ht]
 \caption{Sources, average accretion rate $\dot M_{\rm obs, 18}$
  and accretion period $t_{\rm acc}$ inferred from observations.  
 Coeficients 
 for exponential  ($k_B T_s = a\ e^{-(t-t_0)/\tau}+b$), power-law ($k_BT_s=\alpha(t-t_0)^{\beta_1}$) and 
 broken power-law fits ($k_BT_s=\alpha(t-t_0)^{\beta_1}$,  $k_BT_s=(t-t_b)^{\beta_{2}}$). 
 References: 
[a]~\cite{Cackett2008}, [b]~\cite{Cackett2010} ,  [c]~\cite{Degenaar2011},
[d]~\cite{DiazTrigo2011}, [e]~\cite{Fridriksson2011}, and  [f]~\cite{Degenaar2013}. $\dagger=$ Fit inconsistent with 
 the last observation. $\dagger\dagger=$ Fit considers a constant offset of $b=(77.3\pm 1.0)$\,eV.}             
 \label{table:01}      
 \centering          
 \begin{tabular}{l |c c |c c c c |c c c c }     
 \hline\hline      
Source &
$\dot M_{\rm obs, 18}$
&$t_{\rm acc}$&
\multicolumn{4}{c|}{Exponential fit}&
\multicolumn{4}{c}{Power, broken power-law fits}\\
&(
g\,s$^{-1}$)&(yr)&$a$~(eV)& $\tau$~(d)& $b$~(eV)&$\chi^2$&$\alpha$~(eV)& $\beta_1,\beta_2\times10^{-3}$& $t_{\rm b}$~(d)
&$\chi^2$\\
\hline
\mxb$^{[a]}$&0.07-0.18&2.5&$73^{\pm2}$& $465^{\pm25}$& $54^{\pm2}$&0.8&\multicolumn{4}{c}{---}\\
\hline
\ks$^{[b]}$&0.05-0.3&12.5&$39.8^{\pm2.3}$&$418^{\pm70}$ &$67.7^{\pm1.3}$&2.00$^{\dagger}$&$174.7^{\pm1.3}$& 
$\beta_1=-12.5^{\pm7}$&&0.88\\
\hline
\exo  &0.03&24 &&&& &\multicolumn{4}{c}{}\\
\tiny{\ch}$^{[c]}$&&&$17.2^{\pm1.8}$& $266^{\pm100}$& $106.2^{\pm2.5}$&0.02&\multicolumn{4}{c}{---}\\
\tiny{\sw}$^{[d]}$&&&$13.4^{\pm0.2}$& $192^{\pm10}$& $107.9^{\pm0.2}$&0.34&$135.0^{\pm17.8}$& 
$\beta_1=-30^{\pm30}$& $166^{\pm99}$&0.3\\
&&&&& && &$\beta_2=-60^{\pm20}$&&\\
\tiny{\xmm}$^{[d]}$&&&$17.2^{\pm5.8}$& $133^{\pm88}$& $109.1^{\pm2.2}$&0.06&$141.0^{\pm8.4}$& 
$\beta_1=-40^{\pm10}$& &0.4\\
\tiny{\textit{all data}}$^{[d]}$&&&$14.0^{\pm1.4}$& $220^{\pm65}$& $107.6^{\pm1.5}$&0.39&$135.8^{\pm2.5}$& 
$\beta_1=-35^{\pm3}$& &0.51\\
\hline
\xte$^{[e]}$&1.1&1.6&$36.9^{\pm1.7}$& $133^{+38}_{-25}$& $123.4^{\pm0.9}$&1.07&$168.8^{\pm5.7}$& 
$\beta_1=-30^{\pm13}$& $38^{+24}_{-12}$&0.88\\
& && &&&&&$\beta_2=-69^{\pm4}$&&\\
\hline
\ig$^{[f]}$
&0.2&0.17& 21.6$^{\pm 4}$& 157$^{\pm 62}$& 84.3$^{\pm1.4}$&1.84&147.9$^{\pm12.7}$& 
$\beta_1=-47^{\pm5}$&  & 1.2$^{\dagger\dagger}$\\
 \hline                  
 \end{tabular}
 \end{table*}
\subsection{\mxb}

This source was detected in outburst first in 1976--1979 and again in 1999--2001. Both outbursts 
lasted about $2.5$ years \citep{Lewin1976}.
Its quiescence was monitored by \ch\ and \xmm\ telescopes, the last observation was made by 
\ch\ 11 years after the end of the last outburst (Fig.~\ref{intro}a). 
Assuming an accretion-power luminosity $L=\epsilon\dot{M}c$, with $\epsilon=0.2$, it is possible to estimate
a mean value for the mass-accretion rate $\dot M_{\rm obs, 18}\simeq 0.07$--$0.18$, where 
 $\dot M_{\rm obs, 18}$ is in units of $10^{18}$\,g\,s$^{-1}$ \citep{Galloway2008}.  

The first six observations of this source 
can be interpreted as the crust cooling down to equilibrium with the core.
The evolution of the surface temperature can be well fit with an exponential function 
 \citep{Cackett2008}, which shows 
that the flux and temperature of the last observation 
remained consistent with the previous two \ch\ observations performed 1000 days before. 

Recently, a new \ch\ observation \citep{Cackett2013} showed an unexpected drop in count rate and a change in the spectral shape that cannot 
be explained by continued cooling. Two possible scenarios 
are discussed in that work:  i) it is assumed that the NS temperature remained unchanged and there was an increase in the column density; ii) alternatively, 
the NS surface temperature dropped and the spectrum is now dominated by a power-law component. Future observations of this source are necessary 
to distinguish between these two possibilities (corresponding temperatures are shown as open symbols in Fig.~\ref{intro}a). 

\subsection{\ks}
First detected in 1989 \citep{Sunyaev1989}, the presence of type I x-ray bursts identified
this compact object as a NS. The source was actively accreting for $12.5$~yr 
and the last detection in outburst was in January 2001 with 
a luminosity of  $10^{36}$\,erg\,s$^{-1}$  \citep{Wijnands2001} with an inferred 
$\dot M_{\rm obs, 18}\sim0.1$ \citep{Galloway2008}. 

Its first four years in quiescence were studied by 
\citet{Cackett2006}; they analyzed \xmm\,(XMM) and \ch\,(CXO) observations and fit the data spectrum with an 
absorbed neutron star atmosphere (see Fig.~\ref{intro}b). 
In that work it was not clear whether the source had reached 
 thermal equilibrium with the core or if 
it was still cooling, but the last observation seemed to indicate the first. 
Then, the data were well fit in a first moment by an 
exponential decay to a constant offset (see Table~\ref{table:01}). 

Years later, \citet{Cackett2010} presented a new Chandra observation that shows a decrease in temperature that is 
inconsistent with the previous fit. 
From revising all the \ch\ and \xmm\ data, the authors concluded that the source was still cooling with the temperature 
following a power-law decay (see Tab.~\ref{table:01}). However,  one problem in this analysis is 
that the spectrum may not be purely thermal and some nonthermal contribution may not have been 
detected because of the low number of counts. 
Observations are consistent
with a simple NS atmosphere model, but a low-level (lower than 10$\%$) 
contribution from a power-law cannot be excluded. 

We remark that first observations for \ks\ and  \mxb\  
were 
performed only 25 days after the end of the outburst (similarly as for \exo\ and \ig\ ,  as described below). 
Thus, important information about the first stage of cooling and the physics of the outermost layers is lacking. 

\subsection{\exo}
This source was first detected in 1980 \citep{Parmar1986} at luminosities of ~$\sim10^{36-37}$\,erg\,s$^{-1}$; 
it remained active for more than 24~years. Short X-ray bursts were observed,  and their rise time and duration 
suggested pure helium ignition. The transition from outburst to quiescence 
occurred during 2008 and was monitored by \citet{Degenaar2009,Degenaar2011} and \citet{DiazTrigo2011}. 
They found an uncertain date for the end of the outburst phase, which was poorly constrainted in a period of seven weeks.  
The mass-accretion rate inferred is  $\dot M_{\rm obs, 18}\sim0.03$,  but recent analysis pointed out the 
possibility of this being underestimated by a factor of 5  because of the high inclination 
of the binary system with respect to our line of sight \citep{Degenaar2014nja}.

The quiescent spectrum of \exo\ monitored by \ch\ and \sw\ in the  $19$~months after outburst was described by assuming 
a combination of a NS atmosphere model plus a nonthermal power-law tail (see  Fig.~\ref{intro}c and 
Tab.~\ref{table:01}).  The resulting 
gradual decrease 
in the NS effective temperature (from $\sim124$~eV to $109$~eV) was 
interpreted as crustal cooling by \cite{Degenaar2009}. 
They also observed 
that quiescent light curves present a shift between data thermal 
fluxes (of $\sim6\%$) coming from the two satellites, apparently due to cross-calibration 
problems. 
\cite{DiazTrigo2011} revisited the problem and analyzed \xmm\ data, which are the most 
sensitive observations of the source.
They found that \xmm\ fluxes are compatible 
with \sw, which reaffirms the hypothesis of an offset in the calibration between \ch\ and \sw.

The unabsorbed flux   ($7.7\times10^{-13}$\,erg\,cm$^{-2}$\,s$^{-1}$) detected in April 2010 by \ch\  
is close to the flux measured by the EINSTEIN observatory before the last outburst of the source
($8.4\times10^{-13}$\,erg\,cm$^{-2}$\,s$^{-1}$), supporting the idea that 
the crust has reached thermal equilibrium \citep{Degenaar2011}. Nevertheless, \cite{Parmar1986} reported 
that \exo\ might undergo periods when it is a much fainter X-ray source 
because the accretion disk may completely hide the central emission.
 It is remarkable that there is much less cooling after the end of the outburst than in the other sources.  Recently, a new observation 
of \exo\ was reported \citep{Degenaar2014nja},  showing a lower temperature $\simeq 110$~eV in 2013 and suggesting that the 
crust has not fully cooled yet. This last observation is consistent with the predictions of our simulations (see Sec.~6),  and we 
will include these data in our fits in a future work.

 \subsection{\xte}
The neutron star transient \xte\ was  discovered in 2006 \citep{Remillard2006} and remained in 
an exceptional 
luminous outburst for about $19$~months. The transition from outburst to quiescent emission 
and the first 800 days of the quiescent phase were first monitored by \citet{Fridriksson2010}. 
During most of the quiescent period, the source was followed by 
\ch\ in a campaign consisting of ten observations made between August 2007 and October 2009, 
and, lately, one more in October 2011.
It was also observed three times with \xmm\  in August 2007, September 2007, and March 2009, 
and last data came from  April 2011 taken from \sw\  \citep{Fridriksson2010,Fridriksson2011} (see Fig.~\ref{intro}d). 
Here the luminosity was measured very early in the quiescent phase: 
three data points in the first twenty days. This gives valuable information about the 
the physics of the outer layers of the NS that are directly involved in the cooling  after outbursts.
These early data are a qualitative difference 
 to all other known sources. 
The inferred value for the mass-accretion rate for \xte\ is 
close to the Eddington rate, $\dot M_{\rm obs, 18}=1.1$ \citep{Cackett2010b}. 

Spectra of \xte\ show thermal and nonthermal components which latter is well fit by a 
power-law of index 1-2. The origin of the nonthermal emission 
is poorly understood, but it probably
 originates in magnetospheric activity \citep{Campana1998}. 
The thermal emission in quiescence (see Fig.~\ref{intro}d) 
shows a temperature decrease that is interpreted as
the cooling of the NS crust that was heated up in the accretion phase. 
Nevertheless, some features in the observed luminosity  
indicate that the  crustal cooling may be affected by other 
processes.

First, we note that the effective surface temperature decreases from approximately
$160$~eV to $120$~eV, 
significantly higher temperatures than those inferred for \mxb\ and \ks\ 
(approx. from $120$~eV to $60$~eV).  
The relatively warm surface of \xte\ may be a result of the high
(close to Eddington) accretion rate at which this source has been accumulating matter during most of its active phase. 
Alternatively, it might be due to a higher core temperature (maybe it is a young star?).

Second,  the overall cooling rate seems to be explained by crustal heating, as analyzed in 
\citet{Fridriksson2011} from data from \xmm\ and \ch\, where they found good fits from
considering exponential and broken power-law functions 
with $\chi^2=1.07$ and $\chi^2=0.88$, respectively. 
However, these fits 
do not include the third 
XMM-Newton (XMM-3)  and the fourth  Chandra (CXO-4) observations between $\sim 200-300$ days, 
which show
a considerable increment in thermal and nonthermal spectral components 
(\cite{Fridriksson2011}, see Tab.~\ref{table:01}).

One more drawback is that it  was unclear in \citet{Fridriksson2010}  whether 
the \xte\ crust had already reached a thermal equilibrium with 
the core.  The last \ch\ observation indicated with $80\%$ confidence  that
the surface temperature has decreased, implying that the source is still cooling 
\citep{Fridriksson2011}, which 
is inconsistent with previous fits. 
 
Another challenge for crustal cooling models 
is that the temperatures registered at early times drop on a 
relatively short timescale with an $e$-folding time for the exponential fit of
 $\sim$\,120~days  (compared with $\sim$\,300~days for \mxb\  and $\sim$\,460~days for 
 \ks\  \citep{Cackett2010}), which argues in favor of a highly conductive crust. 
Moreover, the temperature evolution shows a change in the slope at early times of about $80$-$100$~days 
(\cite{Fridriksson2011} obtained even 25-80 days).
This  break in the evolution makes it difficult to reconcile
the initial rapid cooling shown by early observations 
and the much slower decrease from the last data in the same cooling model.

\subsection{\ig}

The transient \ig\ was found in the globular cluster TERZAN by \ch\ telescope in 2003 \citep{Heinke2006}. 
In October 2010 it suddenly entered into an outburst period, increasing its intensity by 
approximately one order of magnitude 
\citep{Bordas2010,Pooley2010}. The source returned to quiescence after about ten weeks \citep{Degenaar2011a}. 
A \ch\ observation 50~days after the end  of the outburst showed that the surface temperature was higher by a factor 4 than 
the base level observed in 2003  and 2009   \citep{Degenaar2011c}.

More recently, \cite{Degenaar2013} reported new \ch/ACIS observations on \ig\ that extend the monitoring to 2.2 years into quiescence. They found that even 
when the thermal flux and NS temperature have decreased, their values still remain well above those measured in the previous accretion phase. They fit these 
last observations with exponential decays and found 
 that when the quiescence base level is fixed to the temperature inferred from the 2003/2009 data, the fit results are poor ($\chi^2\sim3$).  However, this is considerably 
 improved ($\chi^2\sim1.84$) when this parameter varies freely, in which case the base level is $b=(84.3\pm1.4)$~eV,  
 considerably higher than the quiescent level. Because
 this value is close to the obtained from the previous observation in 2013 February, this predicts that the NS crust has nearly leveled off (see solid curve in Fig.~\ref{intro}e).
Nevertheless, the best fit corresponds to a power-law decay with a free base level for which $b=(77.3\pm1.0)$~eV, 
which is significantly lower than the most recent 
observation (see dashed curve Fig.~\ref{intro}e), which indicates a continued cooling of the crust.

\subsection{Brief comparison of the sources}

We can group \mxb\ and \ks\ together since they have similar accretion rates ($\sim0.1\,M_{\rm obs, 18}$),  
evolve in a similar temperature range ($\sim120$\,-\,$60$~eV), and 
are (nearly) leveled off with the core
on comparable timescales ($\sim2000$~days). 
Their data spectra are well fit with an absorbed NS atmosphere, and 
their exponential fits show similar e-folding times ($\sim500$\,-\,$400$~days). Although the data are sparse in time, 
their error bars are relatively low.

In contrast, \exo\ and \xte\ (and partially \ig) present  peculiar characteristics. 
They are warmer than the sources in the 
first group and the  data points show larger error bars (like \exo) or exhibit a much higher variability (as in \xte). 
Their temperatures evolve on a higher range 
($\sim125$\,-\,$110$~eV  for \exo, $\sim170$\,-\,$120$~eV for \xte) than the first two, and 
the  e-folding times are considerably shorter ($\sim$130~days for \xte and  $\sim$220~days for \exo). 
  \exo\ has the lowest accretion rate ($\sim0.01 M_{18}$) but the longest accretion time
($\sim24$~yr),  which can be the origin of its 
high surface temperature. 
More puzzling is 
the small amount of cooling that it shows, its temperature decreases by only $\sim15$~eV from the initial
to the last observation. 
\xte\, instead  has the highest accretion rate, at least ten times higher than the other sources. 
The pronounced break between the early and latest observations slope is not evident 
in the other sources. 
Finally, \ig\ shows similarities to the first group, for instance,  an accretion rate of the same order, but 
has   a considerably shorter outburst time (the shortest  among all sources). Like the second group, it exhibits a short
e-folding time of $\sim$60~days and a temperature drop in the overall cooling of only 
$\sim20$~eV.

\section{Baseline model and thermal evolution}
\label{model}
\subsection{Equation of state}
\label{composition}

At low density we used the BBP \cite{Baym1971} equation of state.
The crust-envelope interface is placed at $(5-6)\times10^{8}$\,g\,cm$^3$
and we continued using the BBP EoS to describe the crust 
up to a density of $1.49\times10^9$\,g\,cm$^{-3}$. To take into account the 
effects of the accretion in the crust composition, we used the EoS 
presented in HZ08 in the range $\rho=(1.49\times10^9-3.5\times10^{13})$\,g\,cm$^{-3}$.
This is a BBP-like EoS, but 
modified by  nonequilibrium nuclear reactions in the crust
(see Sec.~3.3). 
The very high-density region in the inner crust and the core is described  
by a Skyrme-type EoS that considers a nucleon-nucleon SLy effective interaction \cite{Douchin2001}. For this chosen 
EoS the crust-core interface is at 
$0.5~\rho_0$, where $\rho_0$ is the nuclear saturation density. 

Throughout this paper we use two different NS models with masses
$M=1.4$~M$_{\odot}$ and $1.6$~M$_{\odot}$. 
Their properties are listed in Tab.~\ref{table:0}: as the NS mass increases, the crust width decreases, which reduces the  crustal relaxation time as $\Delta R_c^2$.

\begin{table}
 \caption{NS configurations: mass $M$, central density $\rho_0$, stellar radius $R$, 
 surface gravity $g$, and crust width $\Delta R_c$}             
 \label{table:0}      
 \centering          
 \begin{tabular}{c c c c c c c }     
 \hline\hline       
 $M$ &$\rho_{0}$  & $R$          &$g$ & $\Delta R_c$\\
 (M$_{\odot}$)  &($10^{14}$\,g\,cm$^{-3}$) &(km)& (${10^{14}}$\,cm\,s$^{-2}$)& (m)\\
\\
 \hline                    
   1.4& 9.88&11.79 &1.34&944\\  
   1.6 &11.65 &11.61 &1.58&735\\
 \hline                  
 \end{tabular}
 \end{table}

\subsection{Superfluidity}

\begin{figure}
    \centering
    \resizebox{\columnwidth}{!}{\includegraphics[angle=-90]{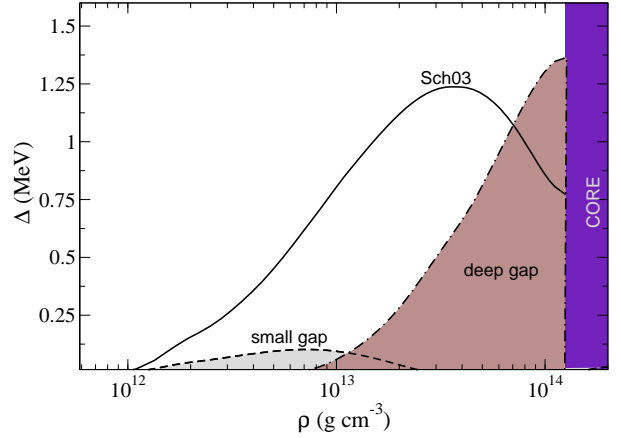}}
       \caption{Energy gap models used as a function of density: 
       Sch03 gap (solid line),  deep gap (dashed dotted line), and  small gap (dashed line).
 }
          \label{fig13}
    \end{figure}

Nucleon pairing does not affect the EoS, but it can play an important role in NS 
cooling since it strongly modifies the specific heat and neutrino emissivities of dense matter.  
Following \cite{Kaminker2001} and \cite{Andersson2005}, 
we used a phenomenological formula for the momentum dependence of the neutron energy gap at zero temperature
given by
\begin{equation}
\Delta(k_{F})=\Delta_0 \frac{(k_{F}-k_0)^2}{(k_{F}-k_0)^2+k_1}\frac{(k_{F}-k_2)^2}
{(k_{F}-k_2)^2+k_3},
\label{Gapsparam}
\end{equation}
where $k_{F}=(3\pi^2n)^{1/3}$ is the Fermi momentum of neutrons and  
the parameters $\Delta_0$ and $k_i$, $i=1..4$ are values fit to microphysical calculations listed 
in Tab.~\ref{Tablegaps}. This expression 
is valid for $k_0<k_{F}<k_2$, with vanishing $\Delta$ outside this range. 
In Fig.~\ref{fig13} we show three different functional forms for the density dependence of the
 neutron superfluidity energy gaps in the NS crust we used throughout: Sch03 gap (from \cite{Schwenk2003}),  deep gap, and  small gap. 

\begin{table}
\begin{minipage}[t]{\columnwidth}
\caption{Parametrization of the energy gaps for neutron superfluidity}
\begin{tabular}{c rrrrr}
 \hline\hline       
Label&$\Delta_0$ ~~ &$k_0$~~~ &$k_1$~~~&$k_2$~~~&$k_3$~~~ \\
     & (MeV) &(fm$^{-1}$)&(fm$^{-1}$)&(fm$^{-1}$)&(fm$^{-1}$)\\
\hline\noalign{\smallskip}
Sch03 & 72.7& 0.1& 6.2 & 1.5 & 2.79\\
deep &4.0& 0.4& 1.5 & 1.65 & 0.05 \\
small & 20.7& 0.1& 6.2 & 1.5 & 2.79 \\
\noalign{\smallskip}
\end{tabular}

References. (a) Sch03 gap; 
(b) deep gap;
(c) small gap
\label{Tablegaps}
\end{minipage}
\end{table}

The corresponding critical temperatures for the $s$-wa\-ve pairing
can be approximately calculated as $T_{\rm crit}=0.56\,\Delta(T=0)$. 
At the relevant densities, the crustal 
temperatures for the five sources are always lower than the corresponding $T_{\rm crit}$,
and neutrons are already in a superfluid state in the inner crust. Unless otherwise stated, 
we considered Sch03 gap in our simulations.

\subsection{Crust composition}
\label{crustcomp}

\begin{figure}
    \centering
  \resizebox{\columnwidth}{!}{\includegraphics[angle=-90]{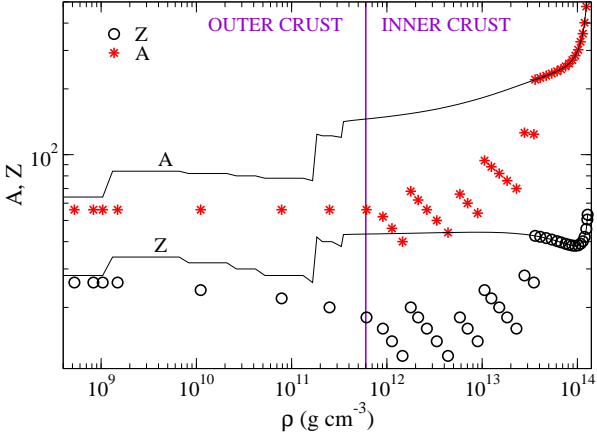}}
       \caption{Accreted crust composition (stars, circles) for $^{56}$Fe burning ashes (HZ08) 
       in comparison to the 
 non-accreted crust (solid lines) (\citep{Douchin2001} above the neutron drip point and BPS \citep{Baym1971} below that density).}
          \label{mdl0a}
    \end{figure}

The crust of an accreting NS can be partially replaced after an accretion period of several years. 
Thus, its composition can be significantly different from that of isolated NSs, see  
Fig.~\ref{mdl0a},  in which  the mass number A (circles) 
and the nuclear charge Z (stars) deviate from the non accreted composition (solid lines) along the NS crust.
   We refer to Sec.~2.1 of HZ08 for details about the capture
rates in different regimes. 

At densities above the neutron drip density, $\rho_{\rm ND}\sim 3 \times 10^{11}$\,g\,cm$^{-3}$,  in the inner crust, 
there are neutron emissions  triggered by electron captures that cause A to decrease.
At higher densities ($\rho > 10^{12}$\,g\,cm$^{-3}$) the mean distance 
between nuclei diminishes and quantum zero-point vibrations 
increase, which leads to pycnonuclear reactions that  result in jumps in A.
In Fig.~\ref{mdl0a} we can observe that the composition  abruptly changes with depth, the jumps
correspond to the location of thresholds for pycnonuclear reactions. 

\subsection{Transport properties and neutrino emission}

The processes that dominate the crust thermal conductivity  (strongly) depend on temperature and density. 
While electron-phonon scattering dominates at low densities in the outer 
crust,  electron-impurity scattering is the most important process at higher densities in the inner crust. 
To calculate these processes, we used the public code of 
A.~Potekhin\footnote{http://www.ioffe.rssi.ru/astro/conduct/condmag.html}.

An important, but uncertain, parameter in calculating  the thermal conductivity is the impurity parameter
\begin{equation}
Q_{\rm imp}= Z_{\rm imp}^2=n^{-1}_{\rm ion}\sum_{\rm i} n_{\rm i}(Z_{\rm i}-\langle Z\rangle)^2,
\end{equation}
which measures the quadratic charge deviation of lattice ions with $n_i$ ion density and charge number $Z_{\rm i}$ 
with respect to the mean value $\langle Z\rangle$ 
and weighted to the mean ion density $n_{\rm ion}$.
High values of this parameter ($Q_{\rm imp}\sim100$) correspond to an amorphous crust 
and a low thermal conductivity. Recent molecular dynamic calculations, however, predict a regular crystalline structure with a 
moderate value of $Q_{\rm imp}$ on the order of unity \citep{Horowitz2007,Horowitz2009} in the outer crust. 
BC09 estimated the value of $Q_{\rm imp}$ by 
fitting the observational data of the sources \ks\ and \mxb\ and also found that $Q_{\rm imp}\sim1-5$.

The crustal specific heat has contributions from the ion lattice, the degenerate electron gas, and the neutron gas in the inner crust that strongly depend on the temperature and density
(see \citet{Aguilera2008} and references therein for the model used here; also  \cite{PageReddy2012} for a detailed discussion). 
Contributions from the neutron gas are suppressed by a  Boltzmann-like factor, controlled by $T/\Delta$ 
(see \cite{Levenfish1994}) when the $T$ falls below $T_{\rm crit}$ for superfluidity. This means that if the neutrons are not superfluid, they  control the specific heat in the inner crust at all $T$. If they were to become superfluid, then this probably is important only in a density region where the suppression is not effective. 
The other two crustal specific heat contributions also vary strongly with $T$ and $\rho$: 
at $T\geq 10^8$~K the ion lattice dominates in most of the crust (and is only overcome by unpaired neutrons).  At lower $T$, the electron contribution is similar to the ionic at $T\simeq 2-3\times10^7$~K  and becomes dominating at $T\leq 10^7$~K, again, at the layers without unpaired neutrons. Superfluid phonons \citep{Aguilera2009} might have a negligible effect or only be relevant in a tinny  region when $T\leq T_{\rm crit}$ and the superfluid phonons velocity approximates the transverse lattice phonon velocity \citep{PageReddy2012}. We neglected the interaction between the ion lattice and the 
neutron gas (pioneering work \cite{Cirigliano2011}, also more recently the entrainment studied in \cite{Chamel2012}), and this might be the main drawback of this approach since it may  influence the thermal evolution of the crust , see \cite{PageReddy2012}. Nevertheless, the results are not conclusive about the strength of 
the coupling between superfluid neutrons and the lattice,  and any type of disorder in the lattice might substantially reduce the effect \citep{Chamel2013}. We plan to investigate this problem in future works.

 We included all relevant neutrino emission processes that 
influence the cooling of the crust (see Tab.~3 
in \citet{Aguilera2008} for a list.)
 At high temperatures ($T\simeq 10^9$~K) 
the dominant process is the plasmon decay. At intermediate values ($T\simeq 5\times 10^8$~K), the  plasmon decay
is only dominant in the outer crust, while electron-nuclei Bremsstrahlung becomes more efficient
in a large part of the crust volume \citep{Yakovlev2001}. We also included the Cooper pair breaking and formation (CPBF) process, 
although it does not affect the thermal evolution of the inner crust.  

\subsection{Thermal evolution}
\label{thermal}
After defining the baseline NS model, 
we followed its thermal evolution by solving the diffusion equation taking into account all 
energy gains and losses:
\begin{equation}
 c_{\rm v} e^{\Phi} \frac{\partial T}{\partial t} + \vec{\nabla} \cdot  
(e^{2\Phi } \vec{F})= 
e^{2 \Phi}( Q_{\nu}+ Q_{\dot m})\,,
\label{eq:eqn1}
\end{equation}
where $c_{\rm v}$ is the specific heat per unit volume, $Q_{\nu}$ denotes the energy loss by neutrino emissions and
$Q_{\dot m}$ considers energy  gains as a consequence of the accretion of matter. 
 As we mentioned in Sec.~\ref{composition}, the deep crustal heating considers that
there are heat sources located 
in the inner crust
as a result  of pycnonuclear 
reactions and electron captures  as well as other less intense sources in the outer crust (HZ08). 
 The metric used is 
$ds^2= -e^{2\Phi}dt^2+ e^{2\Lambda}dr^2 + r^2 d\Omega^2$ and in the diffusion limit  
the heat flux  $\vec F$ is given by the following expression:
\begin{equation}
\vec{F} = -e^{-\Phi } \hat{\kappa} \cdot \vec{\nabla} (e^{\Phi } T)\,
\end{equation}
where $ \hat{\kappa}$ is the thermal conductivity tensor and $ e^{\Phi } T$
is the redshifted temperature. In our one-dimensional treatment the flux is only radial and $\hat \kappa$ becomes a scalar $k$ 
that 
includes contributions of electrons, neutrons, protons, and phonons:  
\begin{equation}
k=k_{\rm e}+k_{\rm n}+k_{\rm p}+k_{\rm ph}
\end{equation}
The electronic term is dominant in the crust, 
while radiative transport is the most important process close to the surface.

The temperature evolution is followed in the 
region that extends from the crust-core interface 
($\rho_{\rm cc}=1.3 \times 10^{14}$\,g\,cm$^{-3}$) 
down to the base of the envelope 
(crust-envelope interface at $\rho_{\rm b}=5.6\times 10^8$\,g\,cm$^{-3}$). 

\subsection{Crustal heating during outbursts: generating the initial thermal profile}

To simulate the accretion phase, we
considered the heat  released per nucleon as a 
function of the density (as in HZ08, Sec.~\ref{crustcomp}).
The integration in Eq.~\ref{eq:eqn1} was iterated 
until the temporal variable equaled the outburst duration.
At this time, the NS crust has reached a thermal profile that depends on the local energy release
per nucleon, the local accretion rate $\dot m$,  the duration of the outburst $t_{\rm acc}$, and the crust microphysics as $c_{\rm v}$. 
Then, the quiescent phase begins and the NS crust starts to cool down from this  
initial thermal profile, 
which corresponds to the conditions at the end of the outburst.

\subsection{Inner boundary: the core}
\label{innerboundary}
The equilibrium temperature of the system 
is set by the core temperature, $T_c$, which mainly depends
on the long-term averaged accretion rate.
We assumed that the recurrence time, that is, the time between two accretion events, is shorter than
the relaxation time of the core ($\sim$10$^3$~yr) and the source has gone through 
several accretion-quiescence cycles, therefore the core has reached thermal equilibrium
and its temperature remains roughly constant. 
Thus,  as an inner boundary condition, we  fixed $T_c$ to a constant value 
taken as a free parameter to fit the observations. If the NS has reached the thermal equilibrium with the core, 
$T_c$ is determined by the last observations. Otherwise, if 
the source is still cooling,  $T_c$ is difficult to infer.

We  checked that assuming a constant $T_{c}$ is a good approximation for quasi-persistent sources 
unless accretion lasts for much longer than $\sim 10$~yr.
In that case, the core could be heated up by 
an inward flux generated by the strong heat deposition over the extended period 
(e.g., for \exo\ if $t_{\rm acc}\sim100$~yr).

\subsection{Outer boundary: the envelope}

To study the thermal evolution of the crust,
the outer boundary condition presents numerical difficulties
since the external layers have a thermal relaxation time ($\sim$1-100~s) 
much shorter than the crustal cooling timescale ($\sim$1000~days). 
Therefore we assumed that the crust is surrounded 
by a fully relaxed envelope and treated the two regions separately. 
 The outer integration limit for the crustal cooling 
is then the bottom of the envelope placed at $\rho_{\rm b}$, with a temperature $T_{\rm b}$,  which is
 influenced by thermonuclear reactions during outburst. In this sense, the initial value of this 
temperature at the beginning of the quiescence phase, $T_{\rm b}^0=T_{\rm b}(t=t_0)$ contains 
relevant information about the heating of the 
envelope during the active phase. In our approach (as in BC09),   we set $T_{\rm b}$ to fit the cooling curves 
to the observational data and 
 leave  the envelope model and  determining $T_{\rm b}$ for a future work.  

\begin{figure}
\vspace{1cm}
   \centering
      {\includegraphics[scale=0.3,angle=-90]{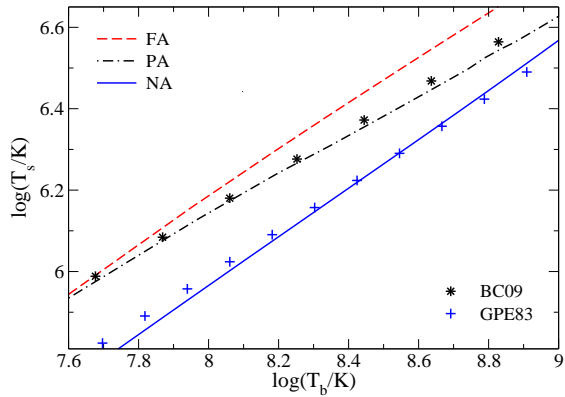}}
      \caption{ $T_{\rm b}$--$T_{\rm s}$ relation used in this work: the partial accreted model (PA, dashed-dotted line) and comparison with the non-accreted Fe envelope (NA, solid line) as in GPE83 (crosses) and the fully accreted (FA, dashed line) in an envelope composed mostly of light elements H and He, as in PCY97.  The relation used in BC09 is shown with stars. }
         \label{env11}
   \end{figure}

The boundary condition for our crustal cooling is 
the  $T_{\rm b}$--$T_{\rm s}$ relation shown in Fig.~\ref{env11}.
  At low $T_{\rm b}$ our partially accreted model (PA) converges to a fully accreted model (FA) composed mainly by H and He, as for example in \cite{Potekhin1997} (hereafter PCY97).  For high $T_{\rm b}$, however, it resembles 
the canonical relation for the non-accreted case (\cite{GPE83}, hereafter GPE83) in a Fe envelope.    
There is an overall agreement between our approach 
and the relation used in BC09 (stars), which  facilitates the comparison of the cooling curves below. 

Linearizing the $T_{\rm b}$--$T_{\rm s}$ relations in a log-log plot for subsequent cooling simulations 
we obtained
 \begin{equation}
 \log (T_{\rm s}/{\rm K})=a+b \log(T_{\rm b}/{\rm K}), 
 \end{equation}
 with slight variations in the coefficients $(a,b)=(2.15,0.49)$ and $(a,b)=(2.20,0.49)$ for a (1.4, 1.6)~M$_\odot$ NS model, respectively.

\section{Revisiting crustal cooling} 
\label{reviewMXB}

We now discuss our results, which we previously compare 
with existing works (see details in Appendix~\ref{App_Comparison}) for testing purposes. 

\subsection{Deep crustal cooling model: testing \mxb}

\begin{figure}
\centering
\subfigure[Temperatures at the crust-envelope interface]
{\includegraphics[scale=0.3,angle=-90]{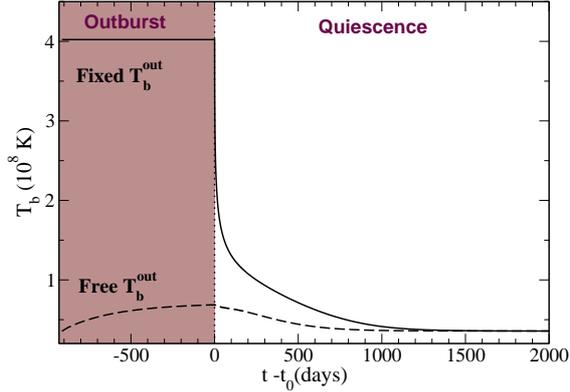}}
\subfigure[Initial thermal profiles ($t=t_0$)]  
{\includegraphics[scale=0.3,angle=-90]{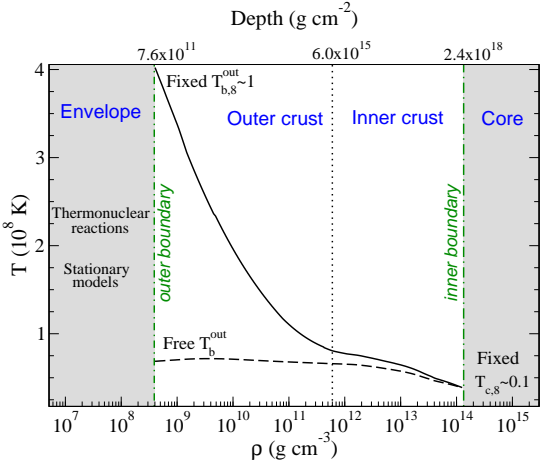}}
\subfigure[Cooling curves]
{\includegraphics[scale=0.3,angle=-90]{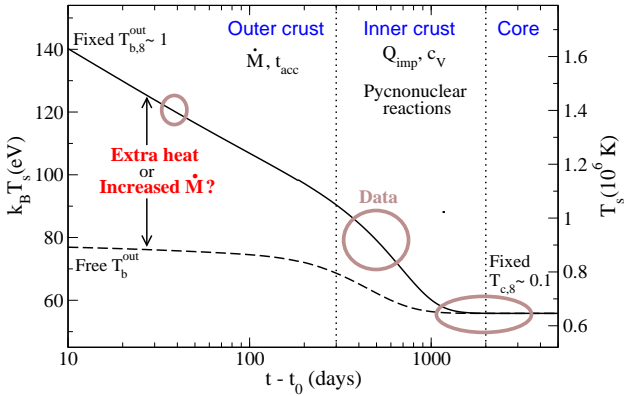}}
\caption{Thermal evolution considering fixed (solid lines) or free (dashed lines) temperature during outburst $T_{\rm b}^{\rm out}$.}
\label{mxbprof}
\end{figure}

We begin with \mxb, which is considered the most standard case.  
We used  a NS with a mass of 1.6~M$_{\odot}$ and radius $R=11.79$~km (Tab.~\ref{table:0}), taking
the impurity parameter  $Q_{\rm imp}$, the accretion rate $\dot M_{18}$, and core temperature $T_{\rm c,8}$ as free parameters. 
The temperature evolution at the outer boundary, \tb, 
during outburst and quiescence,
the corresponding initial thermal profiles, and 
the cooling curve are plotted in Figs.~\ref{mxbprof}$a$,
\ref{mxbprof}$b$, and \ref{mxbprof}$c$, respectively. 

First, we assumed that the temperature at the base of the envelope during outburst, 
$T_{\rm b}^{\rm out}(t\leq t_0)$,
is fixed to  $T_{\rm b,8}^{\rm out}=4.1$ (Fig.~\ref{mxbprof}, solid lines) while 
at the inner boundary the core temperature is kept fixed to $T_{\rm c,8}= 0.29$, 
both values chosen to fit the first and last observations, 
respectively. We set 
$\dot M_{18}=0.1$, $Q_{\rm imp}=4$ and $M=1.6$~M$_{\odot}$, similar to those used in BC09, 
see Appendix~\ref{App_Comparison} for a detailed comparison.
Note that the initial thermal profile suitable 
to explain the data (brown ellipses in Fig.~\ref{mxbprof}$c$) has an inverted temperature gradient 
and hence an inward-directed heat flux. 
As  was pointed out before (BC09, Sht07), the (arbitrary) value of $T_{\rm b}^0$  is crucial 
to explain the early decay. To
illustrate this point, we plot the case when $T_{\rm b,8}^{\rm out}(t\leq t_0)$ is not held fixed,  
but instead evolves freely (dashed curves), controlled only by deep crustal heating (HZ08).  
These curves
fail to explain observations in the early cooling, and
a higher value of $T_{\rm b}^0$  is necessary, as we can see from the cooling curves (Fig.~\ref{mxbprof}$c$). 

At each time, the surface temperature reflects the initial conditions at a particular depth.
Deeper down, the crust did not have time to relax,  and it exhibits roughly
the initial thermal profile. Thus, each depth (or density) corresponds to an evolutionary time.
The early cooling (first 
$\sim 300$~days) is controlled by 
the physics of the outer crust and the initial thermal profile,
which depend strongly on $\dot M$ and on $t_{\rm acc}$. 
The following epoch corresponds to the inner crust thermal relaxation, 
(approx. $\sim (300-1000)$~days) and 
is determined  by  electron-impurity scattering.  
After $\sim 400$~days the imprint of  the superfluid neutron gas
is clear: the temperature fall and the subsequent slope is mostly controlled 
by the strength of the pairing energy gap \citep{Page2013_Proceed} and the possible interaction of the free neutrons with the ion lattice, for example, as a result of entrainment effects 
 \citep{Chamel2013}.

The cooling curve tail reflects the core thermal state (at $t \gtrapprox 1000$ days)
whose temperature remains nearly constant. We checked that the core temperature
is not modified unless the accretion period lasts for much longer than $\simeq$10~yr.

\subsection{Heated-up envelope or incorrect accretion rate?}
\label{Additionalheat}

To show how critical the value of $T_{\rm b}^0$ is for the early decay, 
we also explored the case where $T_{\rm b}^{\rm out}(t\leq t_0)$ evolves freely. 
BC09 estimated that its value cannot rise to $10^8$~K solely by means of deep crustal heating 
(HZ08 sources) and that the required energy release in the outer crust is $\simeq 0.8$\,MeV\,nuc$^{-1}$ for $\dot M_{18}=0.1$, well above that
provided by electron captures (\cite{Gupta2007} and HZ08). Moreover, 
it must be released at a density $\lesssim 3\times 10^{10}$\,gr\,cm$^{-1}$, which 
is again below the density range of electron captures or other known reactions in the outer crust 
(like $^{24}$O burning, \cite{Horowitz2008}). 

\begin{figure}
\centering
\subfigure[Evolution of \tb\ ]
{\includegraphics[scale=0.3,angle=-90]{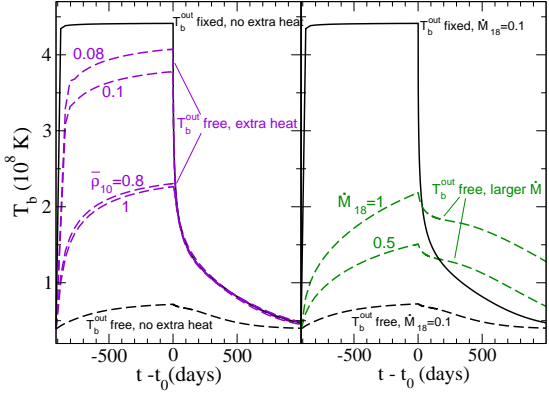}}
\subfigure[Cooling curves  for \mxb\ ]  
{\includegraphics[scale=0.3,angle=-90]{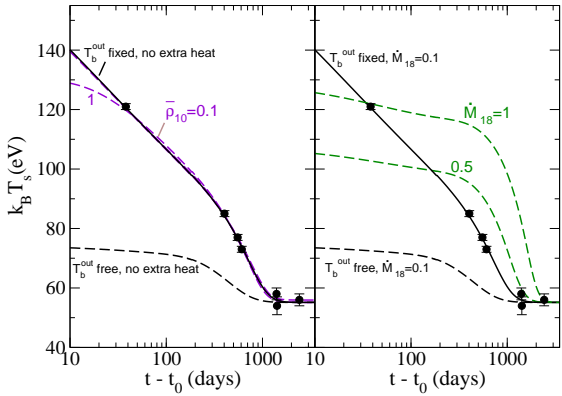}}
\caption{Additional heat or increased $\dot M$? Left panels: 
        dashed curves correspond to $T_{\rm b}^{\rm out}$ free with additional heat sources located
       at $\bar \rho_{10}=0.08,0.1,0.8,1$  with an intensity of $1.6$~MeV~nuc$^{-1}$. 
       Right panel: same as left panel, but instead of additional heat, an increased accretion mass  $\dot M_{18}=0.1,0.5,1$.
       For comparison,  fixed $T_{\rm b}^{\rm out}=4.5$ in solid lines. In all cases $T_{\rm c,8}=0.32$ and $Q_{\rm imp}=3.3-4$.} 
\label{free_fixed_tb}
\end{figure}

The steep fall in the inverted temperature gradient of the initial thermal profile 
is necessary to account for the relatively high temperature 
of the first observation
($T\simeq 120$~eV at 40 days) followed by the moderate value of the second one 
($T\simeq 90$~eV at 300~days), see Fig.~\ref{free_fixed_tb}.  
Our results show that this profile is indeed difficult to achieve unless an additional 
heating source is assumed to be coming from a low-density layer. 
It originate either in the heated-up envelope  
during outburst (that modifies the boundary condition for the cooling 
through the value of $T_{\rm b}^0$) or  in the outer crust at shallow depths. 
This fact was implicitly assumed in BC09 when $T_{\rm b,8}^0$ was fixed to a relatively
high value $\simeq 4$. 
Alternatively,
it has been proposed that \mxb\ 
has been accreting at the Eddington rate $\dot M_{18}\sim 1$,  overestimating
 $\dot M_{\rm obs}$ for \mxb\ by about one order of magnitude (Sht07).  

We simulated these two possibilities: additional shallow heat deposition and increased
accretion rates. 
In the first case we considered the location of additional sources to vary in the range 
 $\bar \rho_{10}\sim(0.08-1)$ 
(where  $\rho_{10}$ is $\rho$ in $10^{10}$\,g\,cm$^{-3}$) with radial width $\Delta r=5$~m and a released energy of 
$1.6$~MeV\,nuc$^{-1}$ while keeping $\dot M_{18}=0.1$ (left panels of Fig.~\ref{free_fixed_tb}). 
We found that the modified $T_{\rm b}$ hardly reproduces the initial steep fall 
of $T_{\rm b,8}^{\rm out}=4$ (solid lines) unless  intense  shallow 
additional sources are present (at $\rho_{10}\lesssim 0.08$). 
If the heat originates in even more external layers, the energy deposited should be lower and 
 the results are much similar to the case with fixed  $T_{\rm b,8}^{\rm out}=4$.  
 This leads to the idea that the heat source is probably located at the top of the outer crust or 
 even in the envelope. 
 
 In the second case, we increased $\dot M_{18}=0.1-1$  (right panels of Fig.~\ref{free_fixed_tb}) without any additional
  heat source. 
 We found that the resulting slope  cannot explain the \mxb\ early data 
(right panels of Fig.~\ref{free_fixed_tb}), which means that it is unlikely that $\dot M_{\rm obs}$ has been underestimated.

We conclude that deep crustal heating by pycnonuclear reactions in the inner crust and e-captures 
in the outer crust is not enough to explain 
the early slope of \mxb\,  and
additional energy from low density regions is needed either from the heated-up envelope during outburst 
or from additional shallow sources in the outer crust.   
Observations shortly after accretion stop are, therefore,  crucial to clarify this point.  

\section{Influence of the thermal state of the envelope and core on the crustal cooling: limiting $T_{\rm b}$ and  $T_{\rm c}$ }

Next we assumed a starting model with fixed $M=1.6$~M$_{\odot}$,  leaving  $T_{\rm c,8}$, and $T_{\rm b}$ as free parameters, usually set in each case to fit the first and the last observation of each source.

\subsection*{\mxb}
\begin{figure}
    \centering
     \resizebox{\columnwidth}{!}{\includegraphics[angle=0]{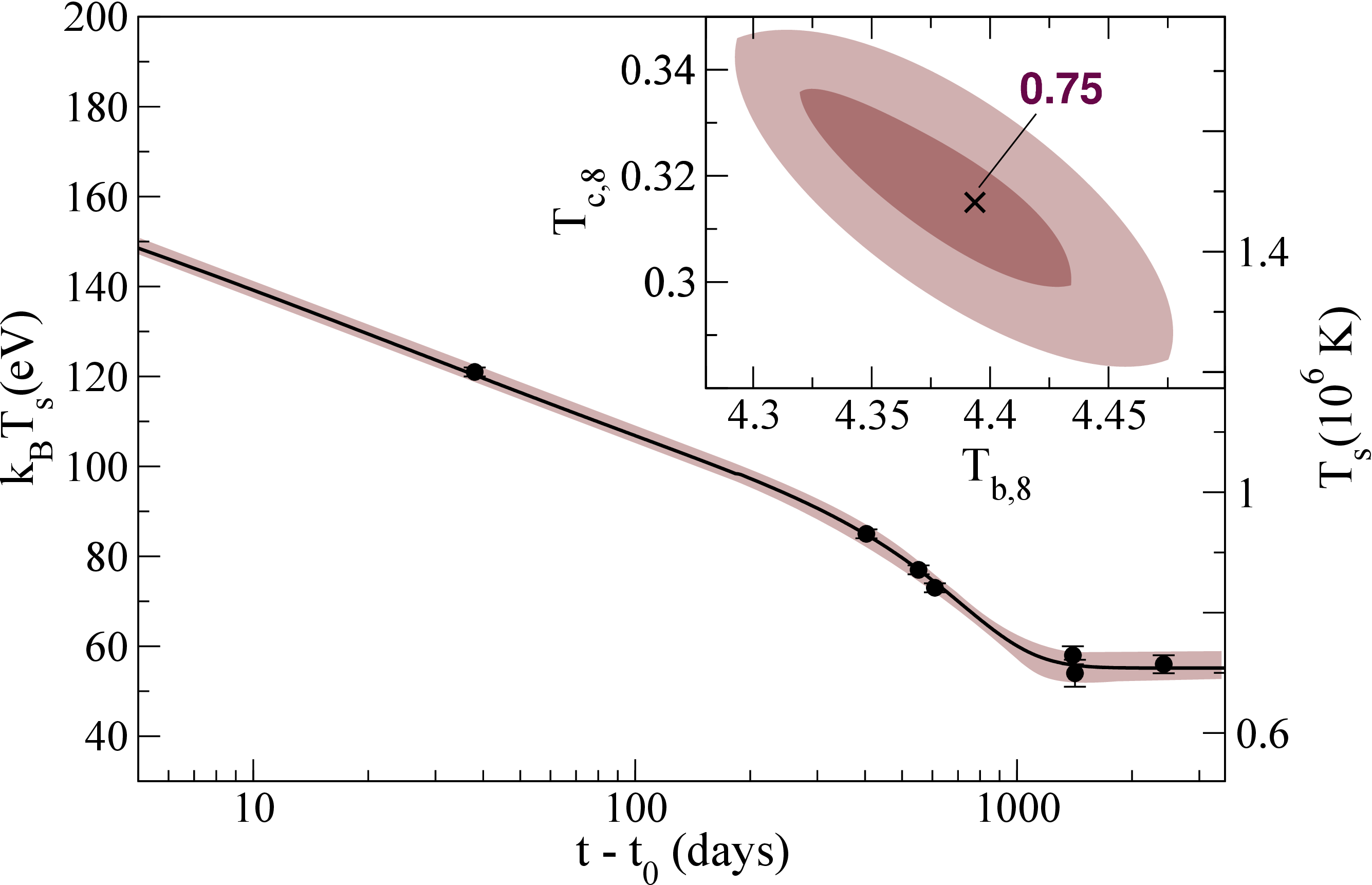}}
       \caption{
 Fits for \mxb. The light
zone corresponds to cooling curves with 
$\chi^2<2$,  the solid line
is our best fit with $\chi^2=0.75$. The inset shows the corresponding
parameter space; the dark zone is for $\chi^2<1$ and the cross the
minimum.  We fix   $M=1.6$~M$_{\odot}$, $Q_{\rm imp}=3.5$, and for the Sch03 gap and vary $\dot{M}_{18}=0.11-0.16$.
 }
          \label{fig10}
    \end{figure}

 We  obtained a thermal evolution that is compatible with a low \q\ value in the crust and found that the source reaches thermal equilibrium 
 in $\sim 1000$~days, which  fully agrees with the simulations of BC09  and  the exponential fits in \citet{Cackett2008}. 
 Cooling curves that fit the observations well (light  regions with  $\chi^2<2$) are shown in Fig.~\ref{fig10}.  We found 
$T_{\rm c,8}=0.28-0.35$ and $T_{\rm b}=4.3-4.6$, the solid curve is the best fit  with $T_{\rm c,8}=0.315$ and $T_{\rm b}=4.39$, as indicated as a cross in the inset.

\subsection*{\exo}

The quiescent luminosity for this source is higher than the
prediction from standard cooling models \citep{Degenaar2011}, and  it has been suggested that 
 residual accretion outside the main accretion period may be responsible for the high temperature 
 (\cite{Brown1998}, \cite{Rutledge2000}, \cite{Colpi2001}). 
Nevertheless, the \xmm\ telescope (which has provided the most sensitive observations)
has not shown features in the light curve that are associated with  residual accretion \citep{DiazTrigo2011}.
Alternatively, it has been   suggested that the core has reached a 
steady state in which the energy radiated 
during quiescence equals the heat released by the reactions taking place during outburst. 
Considering an accretion time of 24~yr and an accretion rate of $\dot M_{18}=0.03$, 
a steady state with such a high temperature would be compatible with a recurrence time of
$\sim$100~yr \citep{Degenaar2011}, a scenario that cannot be ruled out. 

\begin{figure}
   \centering
   \resizebox{\columnwidth}{!}{\includegraphics[angle=0]{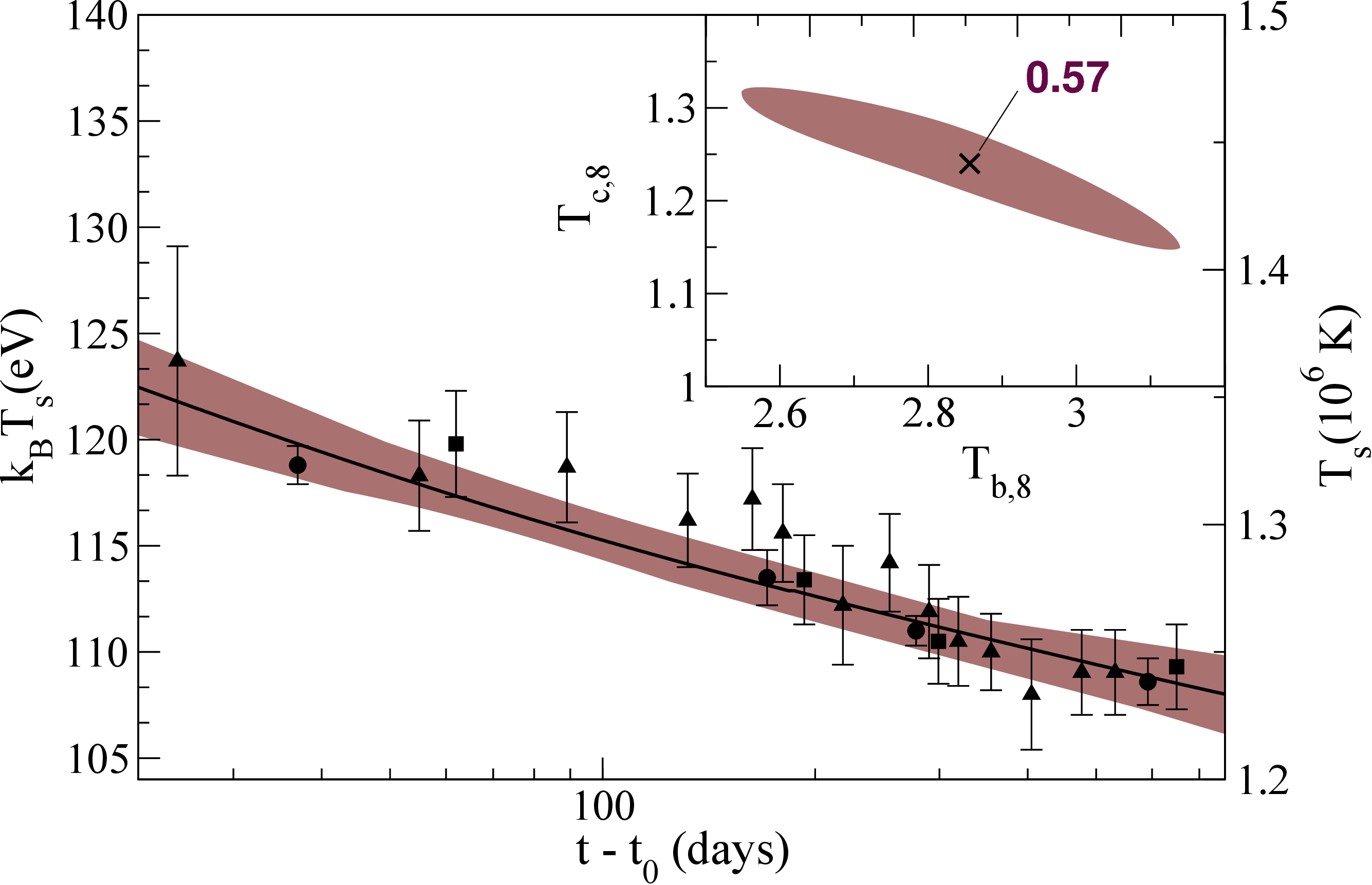}}
      \caption{Same as Fig.~\ref{fig10}, but for \exo\ and only dark
zones with $\chi^2<1$ are shown. The minimum is at $\chi^2=0.57$. We fix
 $M=1.6$~M$_{\odot}$, $Q_{\rm imp}=1$, and the Sch03 gap and vary $\dot{M}_{18}=0.01-0.08$.}
         \label{fig11}
   \end{figure}

Another peculiarity of this source is the low temperature decrease after outburst:
the surface temperature has decreased to a factor of $\sim 0.9$ in 650~days, in comparison 
to $\sim$0.5 for \mxb\ in the same time period.
This is again compatible with a high core temperature and a low accretion rate. 
Another open question is the unknown origin of the power-law component in the spectra.

In spite of such peculiarities, the  quiescent luminosity of \exo\ can also be very well reproduced by a crustal cooling model 
with a rather high core temperature $T_{\rm c,8}\sim$1.25, approx. a factor 3  higher than that of \mxb\ , which
might indicate that \exo\ is a young NS whose core has not yet reached thermal equilibrium. 
The impurity parameter was fixed to $Q_{\rm imp}=1$,
but given that the source is hot, 
the results are quite insensitive to variations of $Q_{\rm imp}$.
 Free parameters varied in the ranges $T_{\rm c,8}=1.15-1.32$ and $T_{\rm b}=2.6-3.1$.
The dark regions in Fig.~\ref{fig11} indicate very good fits with $\chi^2<1$; the best one corresponds to $T_{\rm c,8}=0.57$ and $T_{\rm b}=2.85$. 

Apparently, there is a shift between the \ch\ and \sw\ observations that is maybe 
due to cross-calibration problems between the two satellites \citep{Degenaar2011}.
Even more, \xmm\ and \sw\ fluxes are compatible, 
which also points to an offset in the calibration between \ch\ and \sw\  \citep{DiazTrigo2011}. 
Because of the small error bars, \ch\ data 
allow for a better constraint of \tb, but these data do not provide information about early times. 
Conversely, \sw\ data allow for a better description of the early time,
and \xmm\ data are the most sensitive observations of this source in quiescence 
and, therefore, the most reliable \citep{DiazTrigo2011}. 
We first fit \xmm\ and \sw\ data together to find $T_{\rm c,8}=1.27$, while \ch\ 
data give $T_{\rm c,8}=1.24$. 
Given this tiny difference, we  included all the data in the present analysis\footnote{The observation reported in \cite{Degenaar2014nja} is not included in this work, but we checked that our model can explain it successfully  considering 
core temperatures of $T_{\rm c,8} \simeq 1.4 $.  We plan to publish the results in a future work.}.

\subsection{Summary of the crustal coolers}
\label{fittmxb}

 \begin{figure}
    \centering
       \resizebox{\columnwidth}{!}{\includegraphics[angle=-0]{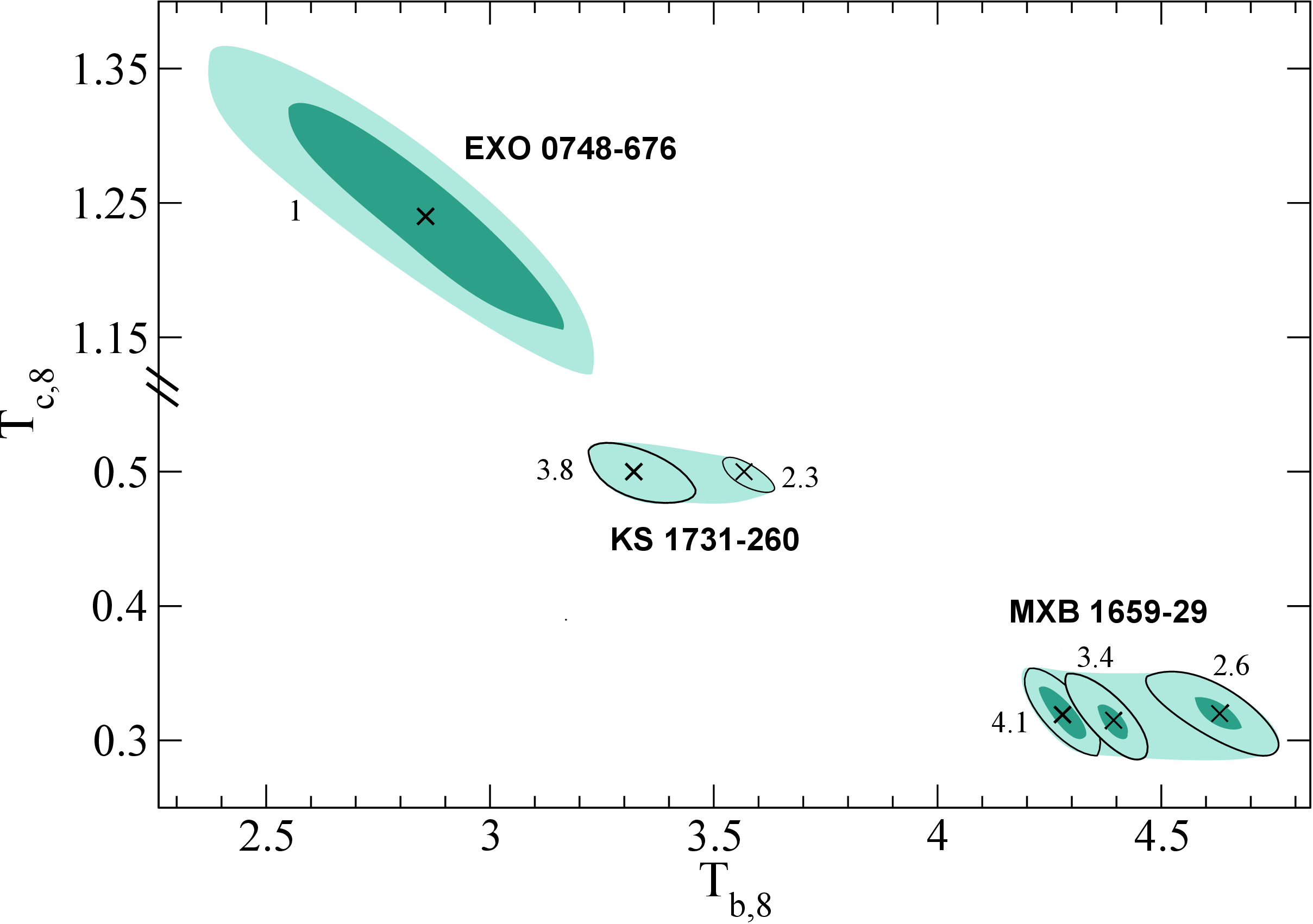}}
       \caption{
Contours for cooling curves for \ks, \mxb, and  \exo\ with $\chi^2<1,2$ (dark and  light zones) by varying $T_{\rm c}$ and $T_{\rm b}$  for $M=1.6$~M$_{\odot}$.  
 Contours with fixed \q\  are shown (values at their sides). 
        }
          \label{fig12}
    \end{figure}

In Fig.~\ref{fig12} we summarized contour plots for the three sources  \mxb, \ks\, and  \exo\ in the  \tc\ - \tb\ pa\-ra\-me\-ter space, 
defined by the conditions $\chi^2<1,2$ (dark and light zones).  We chose a $1.6$~M$_{\odot}$ NS star and a low value of \q ($\lesssim 10$) throughout. We show how the contours for fixed \q\  move in the parameter space.

\section{Toward a model for crustal coolers: constraining the crust microphysics}
\label{constrains}  

This section is devoted to inferring some information about the crust microphysics, and therefore
we mainly focus on the new constraints imposed by the last observation of \ks\ and 
on exploring a model that could simultaneously fit the quiescence emission of 
\mxb, \ks, and \exo. 

\subsection{Is \ks\ still cooling? Constraints on neutron superfluidity energy gaps}
\label{ks} 

The last observations of \ks\ reported by  \cite{Cackett2010} seem to indicate that the 
 source is still cooling, and if this is the case, previous models fail to explain the last temperature drop (Fig.~\ref{intro}b). 
 Indeed,  for our current set of microphysical inputs, none of the curves obtained by varying \q, \tc, or \tb\ 
succeed in explaining  the last observation with $\chi^2 <1$ (no dark zones in Fig.~\ref{fig12}); 
a longer relaxation time with a more effective storage of heat in the crust is needed. 
To achieve this, we explored a neutron energy gap for crust superfluidity with a
relatively low maximum  value, or, alternatively, an energy gap located at deep densities (near the crust-core interface)
such that the resulting suppression of the neutron specific 
heat  and the neutrino emissivity is less efficient.

\begin{figure}
        \centering
         \resizebox{\columnwidth}{!}{\includegraphics[angle=-0]{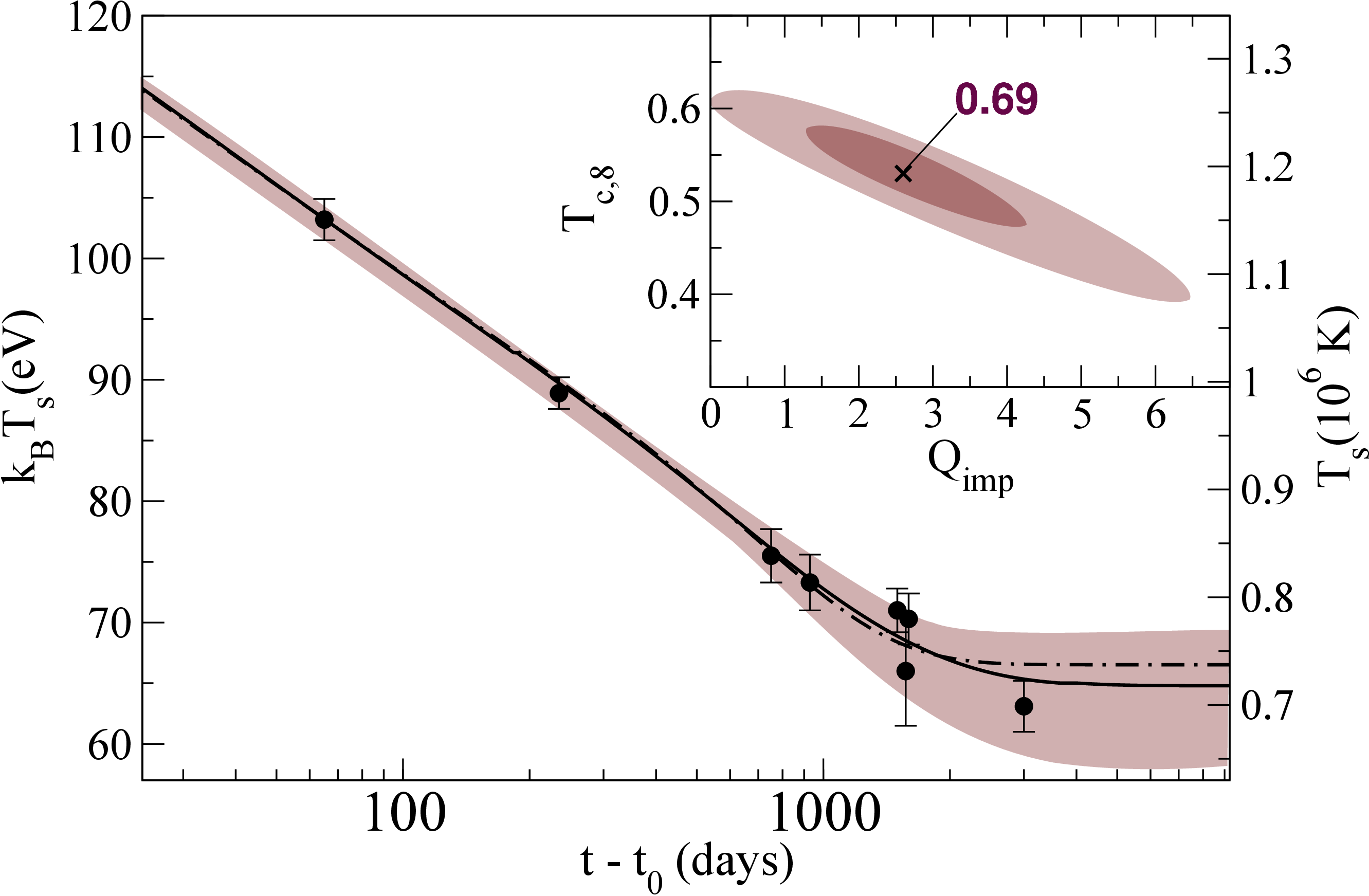}}
           \caption{
           Cooling curves for \ks\ using a deep  gap: 
            the light (dark) brown zones correspond to $\chi^2<2 (<1)$, the solid curve is 
            the best fit ($\chi^2=0.69$) with $T_{\rm b,8}=3.55$. 
            The dotted-dashed curve corresponds to the best fit ($\chi^2=1.14$) with the Sch03 gap and $T_{\rm b,8}=3.52$.
            In all cases, $M=1.6$~M$_{\odot}$, $\dot M_{18}=0.05$, and $Q_{\rm imp}\simeq 4$ are fixed.
            }
              \label{fig14}
        \end{figure}

 In the results presented up to now we used the gap Sch03,  
 which is similar to that used 
 in BC09 or Sht07, which affects 
 the range $\rho\sim (10^{12}-10^{14})$\,g\,cm$^{-3}$ with a maximum value of $\simeq 1$~MeV (Fig.~\ref{fig13}) and 
 results in a thermal evolution that levels off too fast with the core (dotted-dashed line in Fig.~\ref{fig14}. 
 The model we call deep gap 
 (see Fig.~\ref{fig13}) has a maximum 
 located  at $\rho\sim 10^{14}$\,g\,cm$^{-3}$, with an accordingly  longer thermal relaxation time for the crust. 
 Cooling  curves using this  deep gap  can fit the last data point substantially better (brown dark zones in Fig.~\ref{fig14}). 
  A similar effect can be obtained by considering a  small gap with maximum value $\simeq 0.1$~MeV (dashed line in Fig.~\ref{fig13}) 
 or any gap contained  within the colored 
regions in Fig.~\ref{fig13}. 
With these new fits  a lower $T_{\rm c,8} \simeq 0.5$ is reached; future observations are needed to confirm or refute the predicted core temperature.

\subsection{Constraining the impurity parameter}

In this subsection we explore the \q$-$\tc\ parameter space for the three sources \mxb, \ks, and \exo. 

\begin{figure}
\centering
{\includegraphics[angle=90,width=\linewidth]{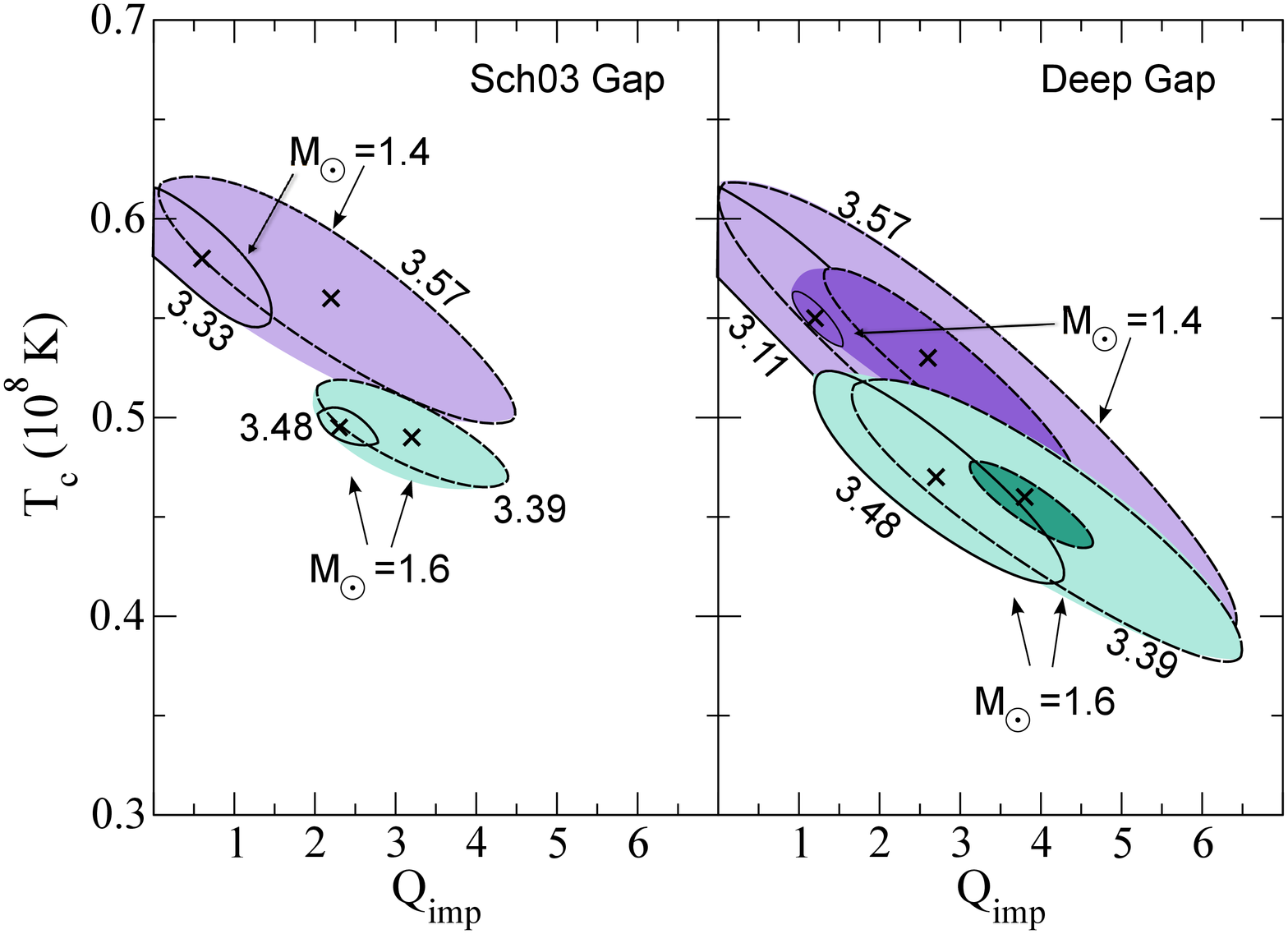}}
\caption{
Best-fit contours for \ks\  comparing Sch03
gap (left panel) and deep gap (right panel). Colored regions
satisfy   $\chi^2<1,2$ (dark and light) considering different
\tb\ ($T_{\rm b,8}$ values at the sides of the contours), accretion rates  (dashed lines for $\dot M_{18}=0.05$, solid lines for $\dot M_{18}=0.1$) and NS
masses (upper, purple contours for $1.4$~M$_{\odot}$ and lower, green contours for $1.6$~M$_{\odot}$).
}
\label{fig15}
\end{figure}

\begin{figure}
\centering
{\includegraphics[width=\linewidth]{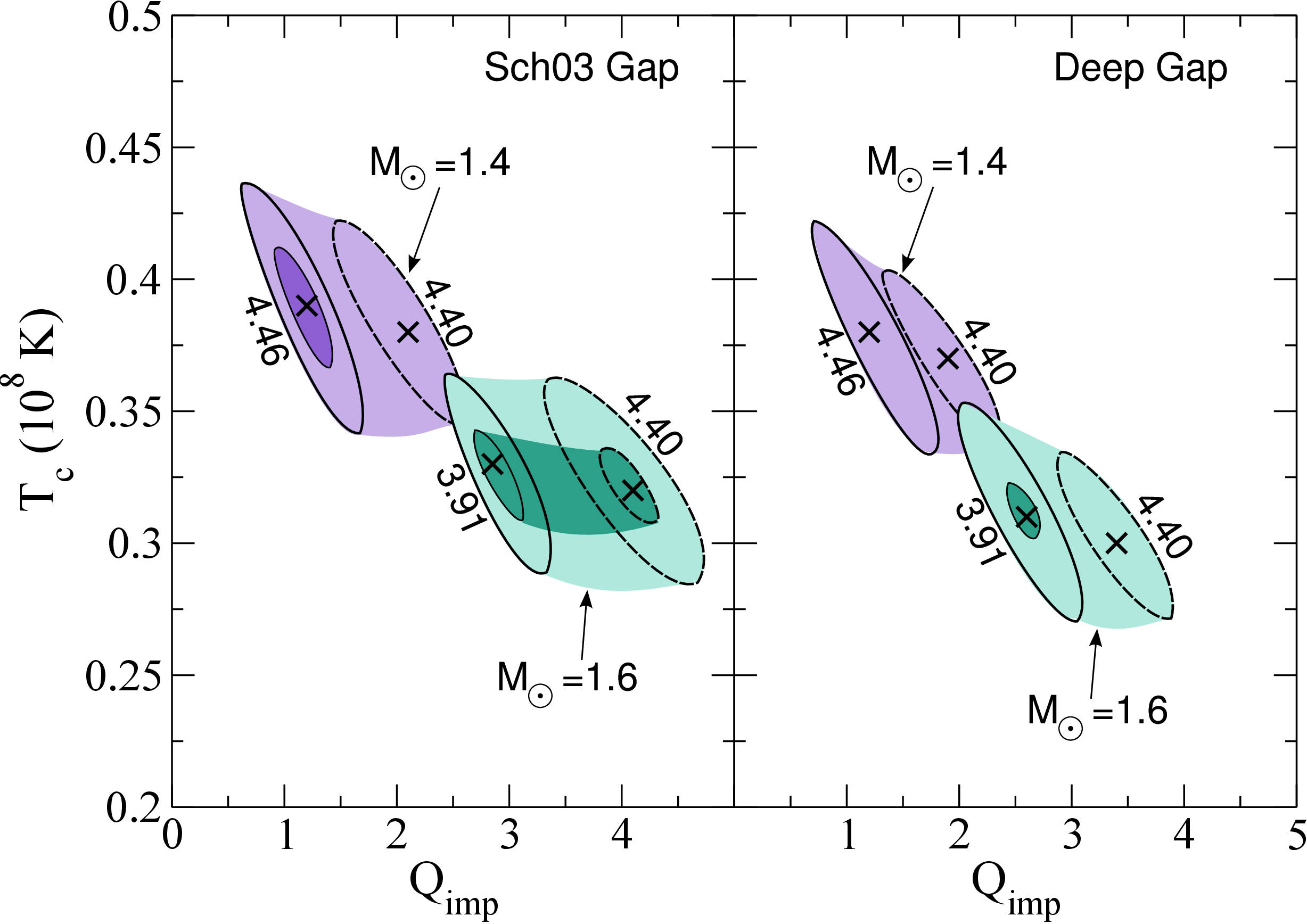}}
\caption{ Idem Fig.~\ref{fig15}, but for \mxb\  and corresponding accretion rates (dashed lines for $\dot M_{18}=0.07$, solid lines for
$\dot M_{18}=0.18$).}
\label{fig16}
\end{figure}

In the left panels of Figs.~\ref{fig15} and \ref{fig16} we show contour levels corresponding to cooling curves that satisfy the conditions $\chi^2<2$ (light regions) and 
$\chi^2<1$ (dark regions) that were obtained for the Sch03 gap and two different masses, $M=1.4, 1.6$~M$_\odot$. 
In right panels, we show the corresponding results for a deep gap. In the different panels we also vary the mass accretion rate $\dot M$ within the observational range 
as much as possible; the ellipses with solid (dashed) contours are calculated for the upper (lower) limit.

 First we note some general  trends in the figures: \\
i) \q\ is correlated with \dm; as \dm\ is increased, the energy released in the inner crust by pycnonuclear reactions  is  increased and overheats the deep layers. To 
balance this effect, \q\ must assume a lower value, which raises the thermal conductivity that favors heat transport to the core.\\
ii) more massive NS has a thinner crust, which reduces the thermal relaxation time, and \q\ shows a shift toward higher values.

In particular, for \ks\ (Fig.~\ref{fig15}), we find that 
  the Sch03 energy gap is unable to fit the data with parameters that satisfy $\chi^2\lesssim 1$, even when varying the NS mass. 
 However, with the deep energy gap the data can be well fit 
  ($\chi^2\lesssim 1$) with $Q_{\rm imp}\sim 3-5$ ($1-4.5$), and $T_{\rm c,8}\sim 0.43-0.48$ ($0.45-0.57$) for $M=1.6$~M$_\odot$ ($1.4$~M$_\odot$).
In contrast, fits for  \mxb\ (Fig.~\ref{fig16}) show that both gaps can fit the data with  $\chi^2<1$.

\begin{figure}[!h]
    \centering
     \resizebox{\columnwidth}{!}{\includegraphics[angle=-0]{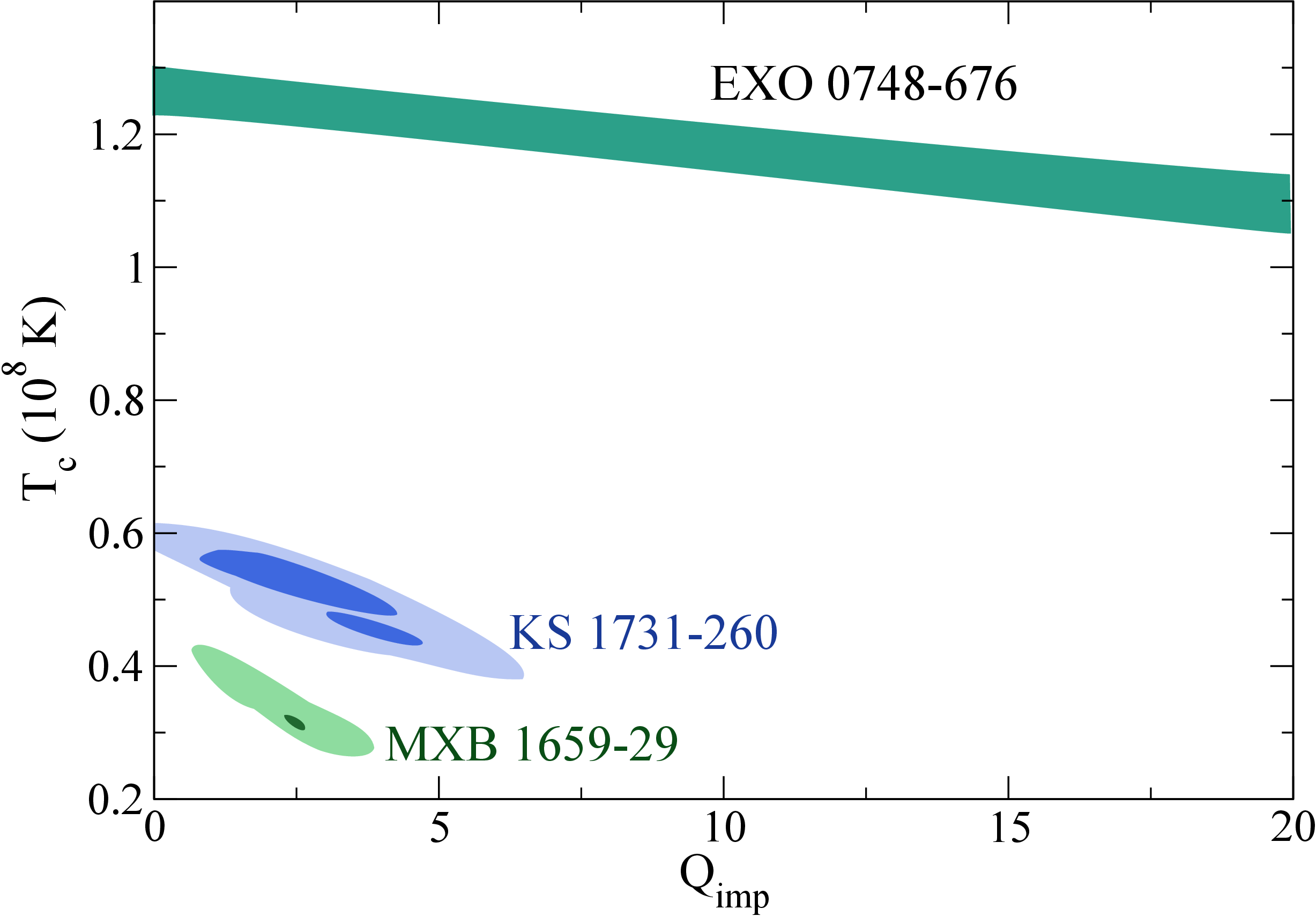}}
       \caption{Constraints for \mxb, \ks, and \exo\ in the $T_{\rm c}$-$Q_{\rm imp}$ parameter space satisfying $\chi^2<1,2$ (dark and light zones) by varying the accretion rate. Deep gap and 1.6$M_\odot$  are assumed.}
          \label{fig17}
    \end{figure}

The comparison of these results with those for  \exo\  is shown in Fig.~\ref{fig17}. Since 
  the last observation of this source was detected at $\sim 600$~days and the
energy gap influences cooling curves only after $\sim 500$~days,  data fits cannot help to 
distinguish between superfluid models. 
Remarkably, because at high temperatures ($T\gtrsim 10^8$~K) the contribution to the thermal conductivity due 
to e-impurities scattering is negligible, cooling curves are barely dependent on \q\ and the 
allowed range for it extends to much higher values than for the other two cases.

\subsection{General trends in the standard crustal coolers}

When comparing results for \mxb, \ks\ and \exo\ sources (summarized in Tab.~\ref{cotas}), we note that while \exo\
 shows a higher equilibrium core temperature, $T_{\rm c,8}\sim 1$, 
  \mxb\ and \ks\ seem to level off at $T_{\rm c,8}\sim$0.3-0.5. We stress that  we obtained good fits ($\chi < 1$) for the 
  three sources with 
  corresponding mass accretion rates compatible with those inferred from observations. 
 Therefore, a model that fits the three sources simultaneously points to the an impurity parameter $Q_{\rm imp}\lesssim 5$. 

 The trend in the fits points to an energy gap for neutron superfluidity with a relatively low maximum value  ($\simeq 0.1$MeV) or peaked at deep densities close to the crust-core interface ($\rho\sim 10^{14}$\,g\,cm$^{-3}$), but this is not conclusive; other processes influencing the contribution of the specific heat and/or the neutrino emissivity may explain the cooling as well.

A  $1.6 M_{\odot}$ star indicates slightly lower values of $T_{\rm c}$ and $T_{\rm b}$ than for $1.4 M_{\odot}$. 
The value of $T_{\rm b,8} \sim 2.8-4.5$ is determined by the thermal state of the envelope  bottom  (at $\rho_{\rm b,8} \sim 6$) at the end of the outburst, and this will set some constraints on the outburst models. 


 \begin{table}[h]
  \caption{ Best fits for \mxb, \ks, and \exo\ with neutron star mass
  $M$, core temperature \tc\ and impurities $Q_{\rm imp}$, considering a deep gap.}             
  \label{cotas}      
  \centering          
  \begin{tabular}{l c c c c c c }     
  \hline\hline       
                       
  Source& M&$\dot M_{18}$&$T_{\rm c,8}$&$T_{\rm b,8}$&$Q_{\rm imp} $& $\chi^2_{\rm min}$\tableBot \tableTop \\ 
   &[M$_{\odot}$] & &&& & \\
      \hline
      
\mxb& 1.4& 0.18  & 0.38 & 4.46& 1.2& 1.18 \\ 
  
  \rowcolor[gray]{0.9}
  & 1.6& 0.18 & 0.31 &3.91 & 2.6& 0.94 \tableBot \tableTop \\
 \hline \hline 
 
\ks  & 1.4&   0.05  & 0.53 &3.39 & 2.6& 0.69 \tableBot \tableTop \\
       
\rowcolor[gray]{0.9}  
    & 1.6 & 0.05 & 0.46& 3.57& 3.8 & 0.87 \tableBot \tableTop \\  
     \hline \hline 
   
      \exo& 1.4& 0.03  & 1.45 & 2.91 & 1.0& 0.58 \tableBot \tableTop \\
        \hline 
            \rowcolor[gray]{0.9}          
          & 1.6 &  0.03 & 1.24& 2.85& 1.0& 0.57 \tableBot \tableTop \\
            
            \hline                  
  \end{tabular}
 \end{table}

Based on this analysis, we call these three sources  standard crustal coolers; despite their differences, their quiescent emission can be explained by means of the heat released by 
pycnonuclear reactions deep in the inner crust, as long as the  NS crust microphysics, models and  boundary conditions (fixing the temperature at the envelope bottom  and in the core) are adjusted. For the other two sources, \xte\ and \ig,  these assumptions 
are not sufficient to account for their quiescent emission and additional heat sources in the outer crust/envelope, residual accretion or new processes affecting the thermal conductivity of the crust have to be assumed, as we discuss next.

\section{Beyond crustal cooling}
\label{beyond}

The peculiar observational data of \xte\ and \ig\ require
models that include additional considerations beyond the deep crustal cooling model controlled by pycnonuclear reactions 
and electron captures in the inner crust. 
In this section we investigate scenarios that could help to understand the quiescent emission for these warm sources:  
an additional heat deposition in the outer crust, a modified heat flow due to a buried magnetic field, 
or residual accretion responsible for the increment/variability in the temperature. 

\subsection{Additional heat in the outer crust?}
\label{addheatsource}

 Previous work on the heat released in the outer crust include 
the report of \citet{Gupta2007}, who  calculated the energy liberated by all thermonuclear reactions  assuming 
a one-component plasma and 
found values $\sim$0.2~MeV\,nuc$^{-1}$. 
Later, \citet{Horowitz2008} calculated reaction rates of $^{24}$O and $^{28}$Ne for a multicomponent 
plasma and found that a composition 
in which $(3-10)\%$ of the ions are $^{24}$O causes reactions that release $0.52$~MeV\,nuc$^{-1}$
 at $\sim10^{11}$\,g\,cm$^{-3}$. This energy could indeed influence the thermal 
 state of the source going into quiescence. 
 
\begin{figure}[!h]
    \centering
  \includegraphics[width=\columnwidth, angle=0]{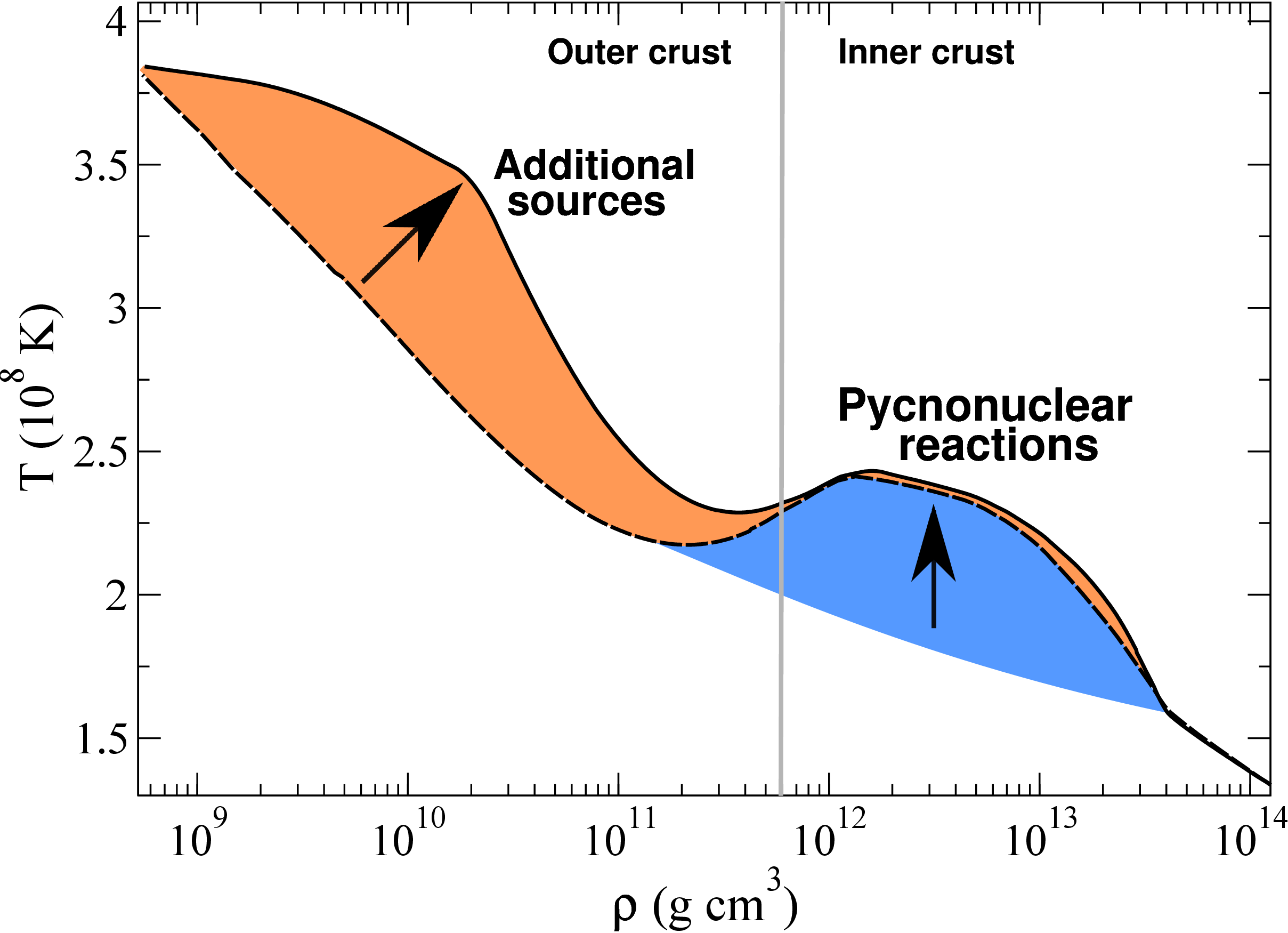}
       \caption{Heat sources that affect the initial thermal profile: pycnonuclear reactions as in HZ08 (dashed line) and additional sources in the outer crust (solid lines). 
       The cooling curves for these thermal profiles fitting \xte\ data are shown in Fig.~\ref{fig20}.}
          \label{xte_addsors}
    \end{figure}

 An initial thermal profile that is peaked in the outer crust  (typically $10^{8-9}$~gr\,cm$^{-3}$, see Fig.~\ref{xte_addsors})
influences the crustal thermal state  
and may be a plausible explanation for the break observed at $\sim 20-150$~days  in \xte\ (Fig.~\ref{sources}d). 
Additional heat sources located in the outer crust that release high enough energy per nucleon 
 could account for these kind of initial profiles (\cite{Fridriksson2010}, BC09, \cite{Degenaar2011a}), as we show next for \xte\ and \ig.

\subsubsection*{\xte}
\label{xtesources}
 
This is the most peculiar source:  
in two observations, XMM-3 and CXO-4, it shows a sudden increase in temperature which so far lacks explanation. 
Without these
two observations,
the exponential fit gives the shortest e-folding time of $\sim100$~days
and the broken power-law fit predicts a break in the slope at about $\sim$25--80~days \citep{Fridriksson2011}
(much earlier than the other sources). 
BC09 suggested that the break is due to the suppression of the specific heat
in the transition from a classical to a quantum crystal. They estimated the time at which the break occurs
(the diffusion time of  the thermal flow from the density at which this transition occurs to the surface) and obtained $\sim 300$~days, 
much longer than expected from the data.

It is also difficult to reconcile  
the early temperature of \xte\ with the latest observations within a cooling model. 
More specifically, we can easily find a set of parameters for the thermal evolution that 
explains the first observation at the third day (COX-1)  and the tail after $\sim 400$~days 
(COX-5 and subsequent), but the problem is to fit the data between $\sim 10$ and $\sim 200$~days
with the same model.

An alternative explanation for the fast initial drop in temperature are additional heat sources in the outer crust that release energy close enough 
to the surface for the heat to be rapidly carried away. After this first stage, 
the temperature evolution should resemble the standard cooling model
without additional heat sources. 
If this is the case, the early data of \xte\ are unique and offer valuable information 
about the depth of the layer where additional heat is released. 
The initial thermal profile is modified by the location of heat sources;
if we consider shallow sources ($\bar{\rho}_{10}\lesssim 10$), the heat accumulated during the accretion stage mostly
diffuses to the surface, keeping the outer crust hot at early times.
Instead, if we consider deep sources, the heat is carried toward the interior and is released by neutrino emission from the core, resulting in lower surface temperatures.

\begin{figure}[!h]
     \centering
      \resizebox{0.95\columnwidth}{!}{\includegraphics{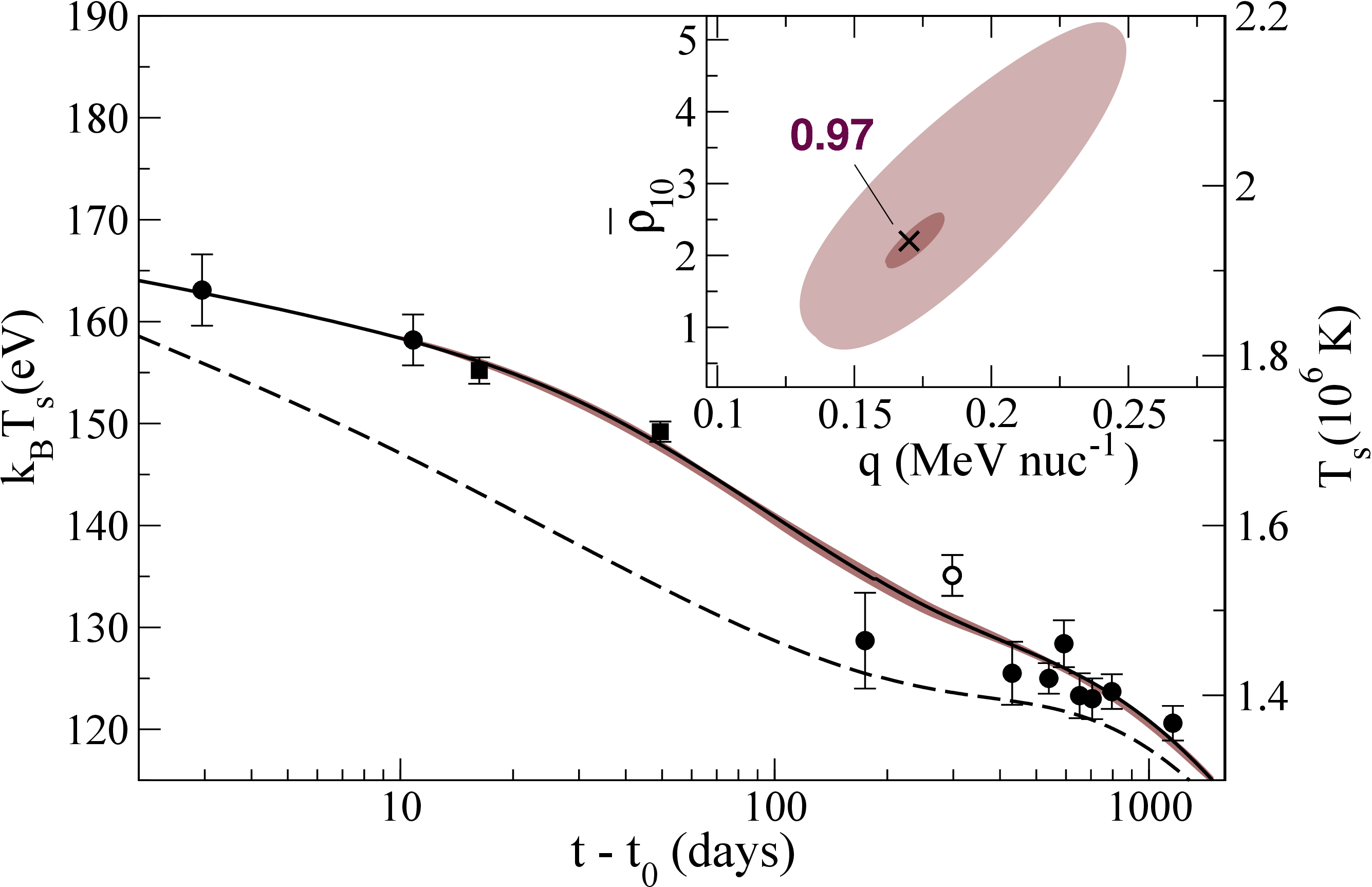}}
        \caption{
        Cooling curves for \xte\ with additional heat sources; best fit with $\bar{\rho}_{10}=2.2$ and 
         $q=0.17$~MeV\,nuc$^{-1}$ ($\chi^2=0.97$, cross in the inset).  The dashed line is the same fit without additional heat sources. 
Fixed parameters: $T_{\rm c,8}=1$,  
$T_{\rm b,8}=3.84$, $\dot M_{18}=1.1$
$Q_{\rm imp}=7$, $\Delta r=20$~m, and $M=1.6$~M$_{\odot}$. 
                }
           \label{fig20}
     \end{figure}

We performed simulations considering that the additional heat is located in a shell characterized by  
the mean density, $\bar{\rho}$, at which the energy is deposited and its radial width, $\Delta r$. 
The results presented below are weakly sensitive to  $\Delta r$ in the range $(1-50)$m, therefore we kept $\Delta r=20$~m fixed and
 chose  a \tb\ value compatible with observations
for an accretion rate of $M_{\rm 18}=1.1$. 
By varying $T_{\rm c,8}=1.0-1.2$ to adjust 
 the first data  (COX-1)  and the tail  (COX-5 and subsequents), we found that the best fit to the 
 intermediate observations gives $\bar{\rho}_{10}=2.2$ and $q=0.17$~MeV\,nuc$^{-1}$ (solid curve in Fig.~\ref{fig20}).

 Compared with the calculations of \citet{Horowitz2008}  for the $^{24}$O$+^{24}$O 
 reaction\footnote{We infer $\bar\rho_{10}\sim10$ and $q=\sim0.1$~MeV\,nuc$^{-1}$, assuming that only $10\%$ of $^{24}$O was burned.}, 
 our simulations predict a  value of $\bar \rho$ for the heat deposition one order of magnitude lower. 
 This difference might be reduced if the effect of neutron skin dynamic is included in the approach, 
 which may result in a significant enhancement of the reaction rate and, hence, a lowering of the corresponding density for the location of the sources.

\subsubsection*{\ig}

 This is the first regular transient with a short active phase of $\sim$  weeks/months  showing evidence of crustal cooling. 
 It is remarkable that having been accreting for a much shorter period
  than the quasi-persistent sources,  its thermal flux  remains, after 2.2~years, still well above 
  the quiescent emission value detected before outburst. 
 For such a short active phase,  it would be expected that the crust reaches a lower temperature.
 
\begin{figure}
     \centering
     \resizebox{0.95\columnwidth}{!}{\includegraphics[angle=-0]{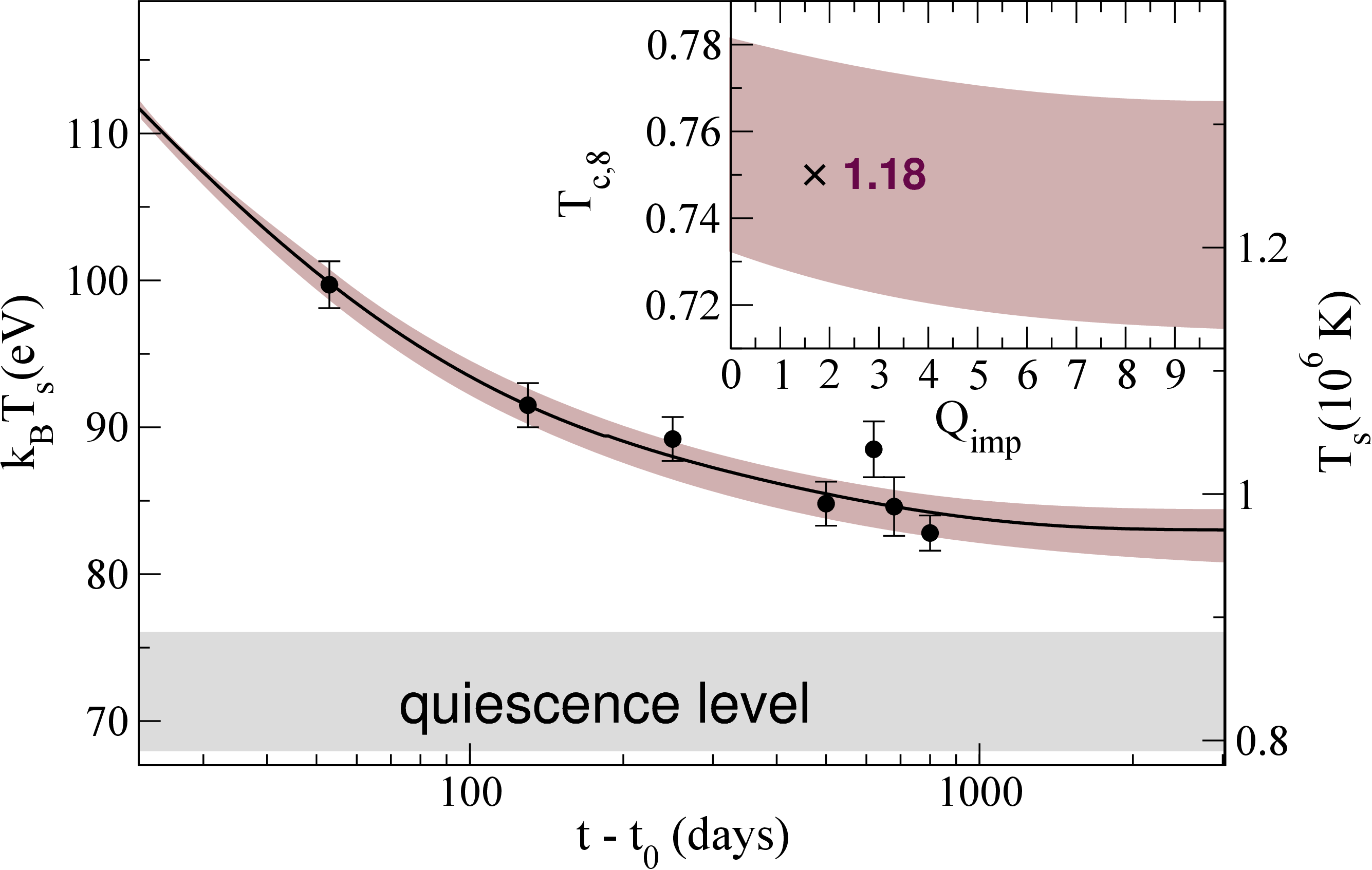}}
        \caption{
        Best cooling curves for \ig; the brown zone corresponds to $\chi^2<2$,
         and the solid line is the best fit with 
         $\chi^2=1.18$  (cross in the inset). We fix $\dot M_{18}=0.2$, $T_{\rm b,8}=4.06$, and the Sch03 gap.}
        \label{fig21}
     \end{figure}

The information on the previous quiescent equilibrium level  imposes a constraint on \tc. 
If we leave \tc\ as a free parameter in our fits,  
we find that the NS crust levels off after $\sim2000$~days at a temperature of $T_{\rm c,8}=0.75$, which is far above than 
the quiescent level before outburst $\sim (0.44-0.55) \times 10^8$~K (see solid curve in Fig.~\ref{fig21}), 
in agreement with the results of \citet{Degenaar2013}. 
 
\begin{figure}
     \centering
     \resizebox{0.95\columnwidth}{!}{\includegraphics[angle=-0]{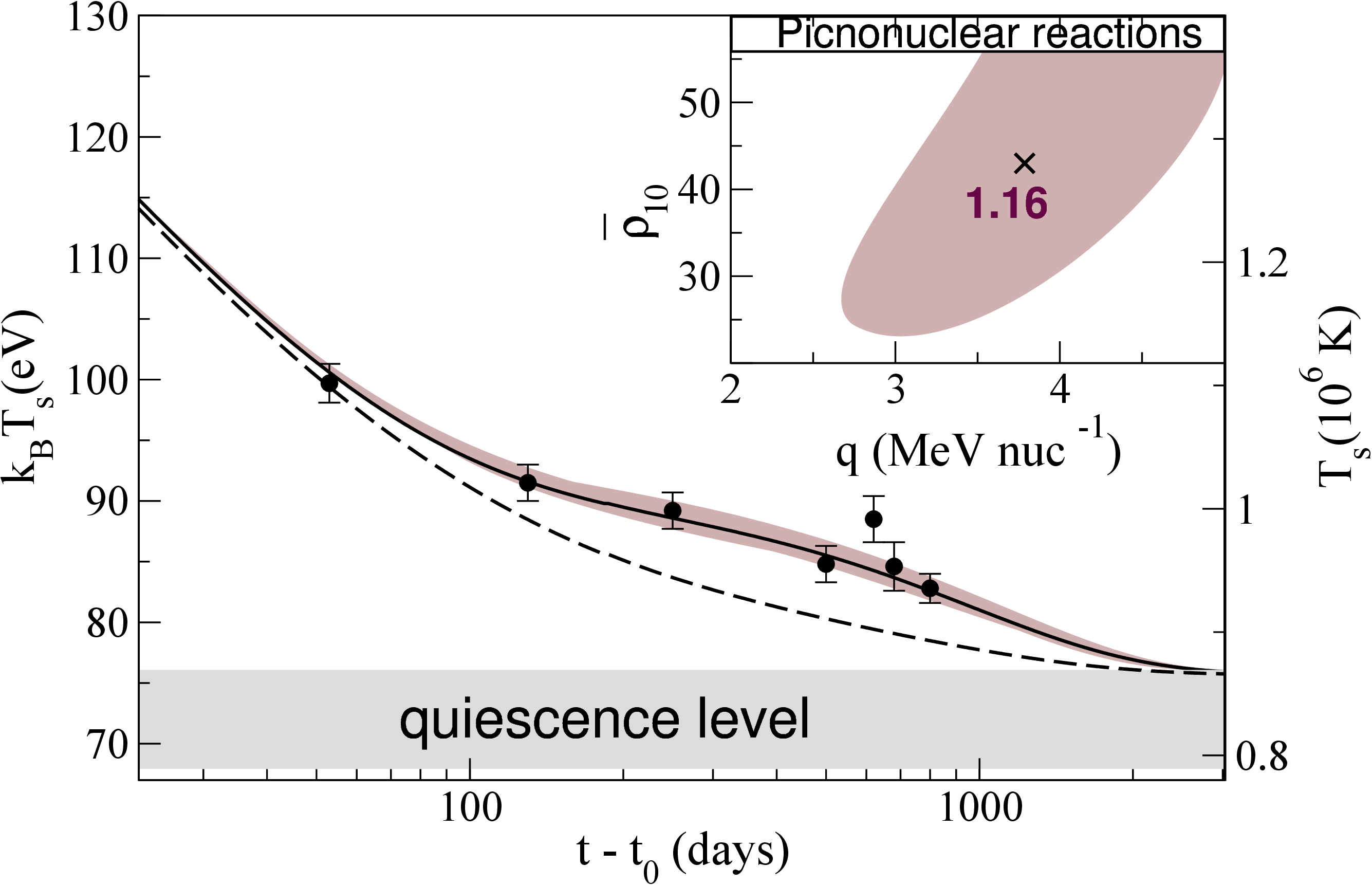}}
        \caption{  
        Idem Fig.~\ref{fig21}, but with additional heat sources optimized with respect to $\bar{\rho}_{10}$ and $q$. 
        The brown region and light contours in the inset correspond to $\chi^2<2$. 
        Solid line is the minimum  $\chi^2=1.16$ (cross in the inset). 
         We fix $T_{\rm c,8}=0.62$, $T_{\rm b,8}=4.62$,  $\dot M_{18}=0.2$, Q$_{\rm imp}=7$, and the Sch03 gap. 
         The dashed line fixes $T_c$ to the quiescent level without additional heat sources. 
          }
        \label{fig22}
     \end{figure}

Alternatively,  if we fix \tc\ to the value in the previous quiescence period, all cooling curves underestimate the late-time temperatures
(dashed curve in Fig.~\ref{fig22}). One possible solution is again an additional heat source
\citep{Degenaar2011a,Degenaar2011b}; results are plotted in Fig.~\ref{fig22}. Models satisfying  the condition $\chi^2<2$ are shown as 
light brown region, and the best fit corresponds to $q=3.8$~MeV\,nuc$^{-1}$  and $\bar \rho_{10}=43$  (solid curve).
Comparing these results with those found previously for \xte, we note that these heat sources are extremely intense and deeply located,
and its origin can hardly be determined. 

We conclude that there are two different possibilities that can explain the observations of \ig. First, 
standard cooling (without additional heat sources) but with an equilibrium temperature well above  the value measured in the previous quiescent phase. 
This can be a consequence of a change in the \tbts\ relation with respect to the previous quiescent phase 
(because of a change in the envelope composition during the accretion phase), which could set a higher observed equilibrium level for the same interior temperature \citep{Degenaar2013}. 
Second, it is also possible to fit the data by fixing \tc\ in a value compatible with the quiescent band, but then it is necessary to consider very intense additional heat sources whose origin is unclear.
Future monitoring of \ig\ will determine if the source has leveled off, favoring the first scenario, or whether it is still cooling, which would indicate non-standard cooling.

 \begin{figure}
     \centering
        \resizebox{0.95\columnwidth}{!}{\includegraphics[angle=-90]{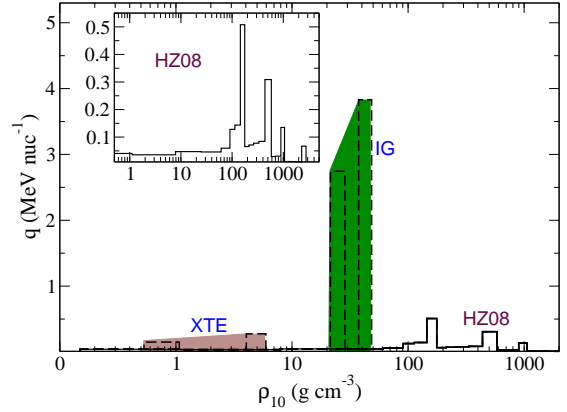}}
        \caption{ Additional heat sources distribution and intensity for \xte\ (brown region), 
        and \ig\ (green region). 
        Solid lines are HZ08 sources (inset). }
        \label{fig23}
     \end{figure}
     
To summarize, in  Fig.~\ref{fig23} we compare the additional heat sources needed to explain the quiescent emission of \xte\ and \ig\  with the
theoretical calculations of the heat deposited by crustal heating from HZ08.    
Colored bands illustrate how the source intensity is modified when they are located at different depths. 
The inset shows  the HZ08 results in more detail.

\subsection{Buried magnetic field}
\label{isolation}

Another possibility that might explain a warmer outer crust at early times is a 
low conductivity layer between outer and  inner crust. This can be the result of a  buried magnetic field, 
as suggested by \citet{PayneMelatos2004}, if  the 
magnetic field lines are pushed into the crust and concentrate in a thin shell during the accretion period. The thermal conductivity 
would be highly reduced in the thin layer, and would act as a thermal insulator between outer and inner crust. 
The cooling curves will be affected by the suppressed thermal conduction, resulting in an accelerated 
cooling at early times (released of the heat deposited in the outer crust) followed by a slower temperature decrease. 

\begin{figure}
    \centering
     \resizebox{0.95\columnwidth}{!}{\includegraphics[angle=-0]{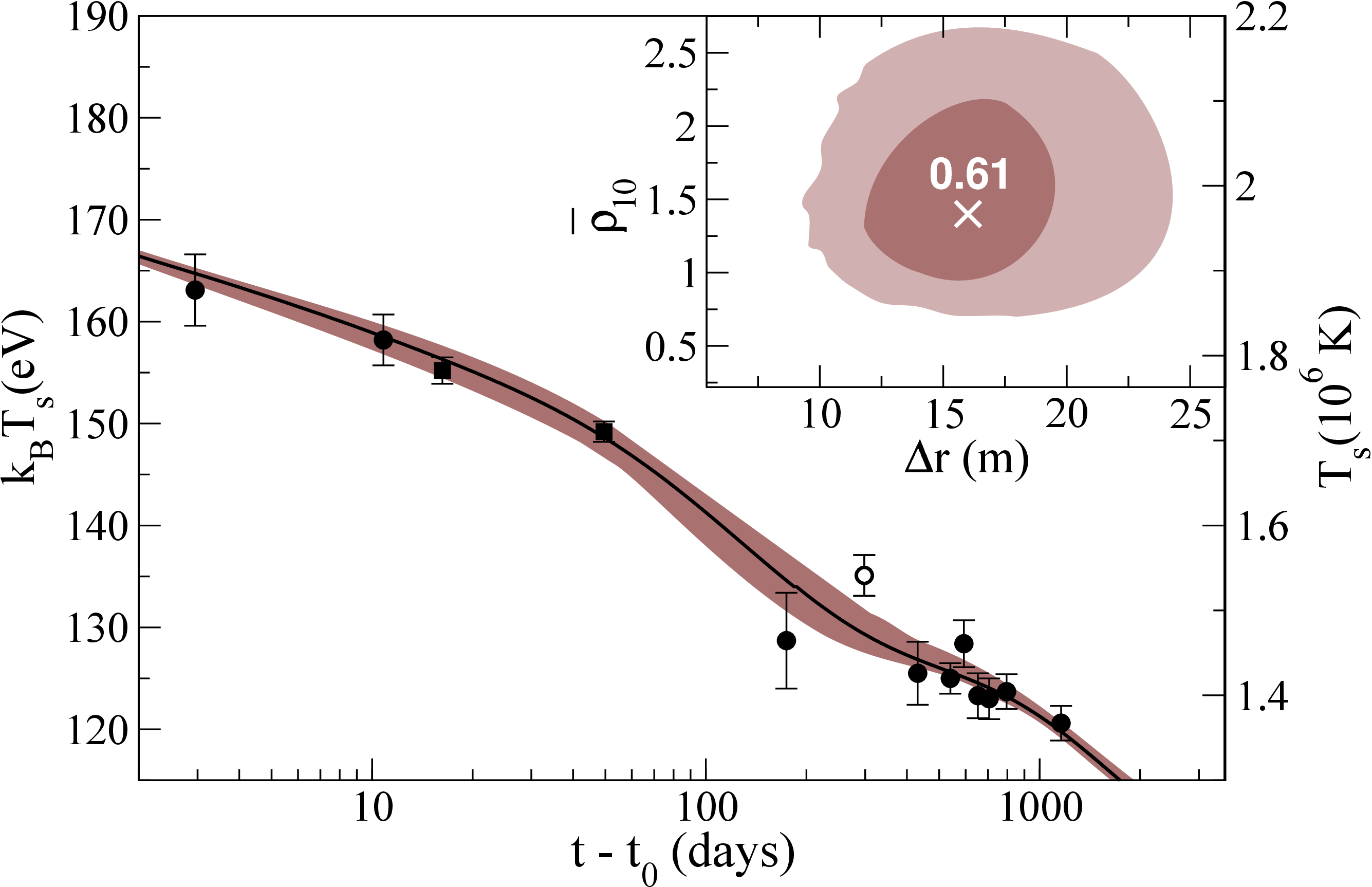}}
       \caption{ 
       Cooling curves obtained by suppressing the electron thermal 
conductivity by $R_{\rm sup}=0.1$. 
The brown dark (light) regions correspond to $\chi^2<1(<2)$, the solid line to $\chi^2=0.61$. 
We fix
$T_{\rm c,8}=1.35-1.40$, $\dot M=1.1$, $T_{\rm b,8}=4.05$, Q$_{\rm imp}=1$, and the Sch03 gap. The inset shows $\chi^2<2$ contours in parameter space by varying $T_c$.}
          \label{xte12}
    \end{figure}

To test this hypothesis, we suppressed the electronic thermal conductivity with a factor
$R_{\rm sup}=0.1$ in a layer characterized by its radial width, $\Delta r$, and the mean density 
$\bar\rho$ at which the suppression occurs, fixing the accretion rate to the observational value, $\dot{M}_{18}=1.1$ and  $T_{\rm b,8}=4.05$.
Results show that 
the parameter range compatible with the observations is
$\bar\rho_{10}=(0.8-2.2)$ and $\Delta r=(11.9-19.5)$\,m (Fig.~\ref{xte12}) and with a variation of $T_{\rm c,8}=1.35-1.40$. The minimum is located at  $\bar\rho_{10}=1.3$ and 
$\Delta r=15.8$\,m (with $\chi^2<1$). 
To  analyze this more  exhaustively  it is necessary to study the influence of the magnetic field geometry on the results
in a 2D model, which is far beyond the scope of this work.

 \subsection{Residual accretion in \xte ?}
 \label{Appendix_ResAccretion}
\begin{figure}
\centering
\subfigure[  Evolution of  $T_{\rm b}$ assumed during residual accretion episodes A and B (dashed and dotted lines, respectively).]
{\includegraphics[scale=0.27,angle=-90]{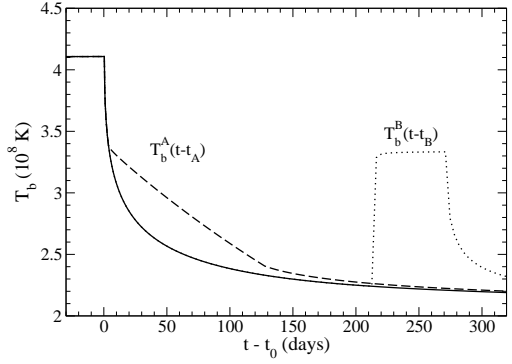}}
\subfigure[Cooling curves from fitting filled symbol data without residual accretion: solid line is the best fit with $\chi^2=0.53$ and brown band has $\chi^2 < 1$
 (with $\dot M_{18}=0.7-1.3$ and $T_{\rm c,8}=1.3-1.8$).
The dashed and dotted lines ($\chi^2=0.55,0.63$) include residual accretion through $T_{\rm b}^{A,B}(t)$, respectively.]  
{\includegraphics[scale=0.27,angle=-90]{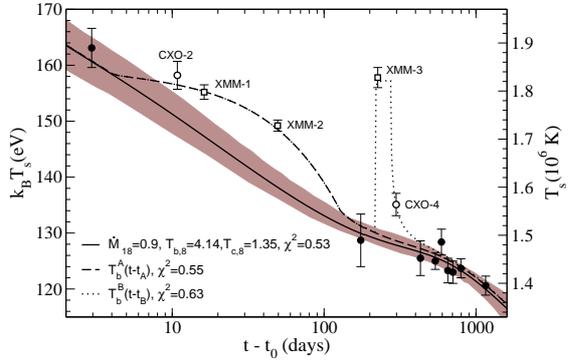}}
\caption{Residual accretion episodes during quiescence in \xte. }
\label{fig19}
\end{figure}

As an alternative scenario, we speculate that some data points in the emission of \xte\ may exhibit a higher temperature as a result of
 residual accretion episodes than a  baseline standard cooling. 
Then we added two residual accretion periods: in the first $\sim 150$~days 
the period $A$ coincident with CXO-2, XMM-1 and XMM2 
and later, at about  $200$~days, with a duration of $\sim 60$~days,
the period $B$ in correlation 
with XMM-3 and CXO-4. 
 We mimicked the change of the accretion mass rate with a 
variation of the  temperature at the base of the envelope $T_{\rm b}$ for the periods $A$ and $B$ 
as  shown in Fig.~\ref{fig19}a. 
We took 
exponential decay functions $T^{A,B}_{\rm b}(t)=T_{\rm b}^{0}e^{-(t-t_{A,B})/\mu}$ with 
$T_{\rm b,8}^{0}=4.1$, $\mu=475$~days, $t_A=t_0$ and $t_B=213$~days for the accretion periods 
$A$ and $B$  (Fig.~\ref{fig19}a). 
The corresponding
cooling curves with residual accretion included are shown in Fig.~\ref{fig19}b. A NS with $M=1.6$~M$_{\odot}$ was assumed. 
The brown region demarcates the curves that fit 
filled points with $\chi^2<1$.
The solid line is the best fit without accretion after $t_0$; it predicts that the source
is still cooling down. The dashed line is the temperature evolution including 
$T_{\rm b}^{A}(t)$ and fits the filled symbols plus CXO-2, XMM-1, and XMM2.
The dotted line fits all the points, and  also XMM-3 and CXO-4 assuming an accretion rate of
$T_{\rm b}^B(t)$. Roughly estimated, we need a mean accretion rate of  about $20\,\%$ of the value of $\dot{M}$ 
during outburst to account for all observations. 

Even though our results show that residual accretion can explain  the CXO-2, XMM-1, and XMM-2  observations, \citet{Fridriksson2011} stated that
the thermal component outside flares (XMM-3 and CXO-4 observations) is probably not caused by ongoing low-level accretion. 
This is because the temperature evolution throughout quiescence (excluding flares) presents an smooth and monotonic decrease; if accretion contributed significantly to the thermal emission 
much more irregular variability would be observed in this component.  
Moreover, no correlation between thermal and non thermal fluxes outside flares has been observed, whereas both rise together during flares. Nevertheless,  residual accretion 
outside flares is a possibility that cannot be conclusively ruled out.  

\section{Summary}
\label{conclusions}

We have presented detailed numerical models describing the thermal relaxation of the crust following
long accretion periods. This was motivated by the increasing number of observations of  \mxb, \ks, \exo, \xte, and \ig. 
Our main results are summarized as follows: 

\begin{enumerate}

\item First, we checked by fitting \mxb\ observations that the energy released by pycnonuclear reactions ($\sim$0.05~MeV\,nuc$^{-1}$) does not seem to be enough to explain the high initial temperature ($\sim 10^{8}$~K), which confirms the results of BC09. Therefore, to explain the early slope of the cooling curves of \mxb, it its necessary to consider an additional inward-directed heat flux that originates in outer layers. 

\item  We solved  the thermal evolution of the neutron star crust as it cools down 
taking into account deep crustal heating and we successfully fit \mxb, \ks, and \exo\ observations by adjusting the neutron star microphysics. We also obtained, in agreement with previous works, that the impurity content $Q_{\rm imp}$ has a low value 
($\lesssim 5$). 

\item We also studied the influence of neutron superfluidity on the results. \mxb\ and \exo\ can be modeled 
with the same microphysics as in BC09. However, \ks\ imposes an additional constraint on the neutron energy gap. The last observation suggests a longer relaxation time, which is compatible with
an energy gap for neutron superfluidity that has a low value ($\lesssim 0.1$~MeV) 
or is peaked at a relatively high density, deep in the inner crust ($\rho\sim10^{14}$\,g\,cm$^{-3}$), although this is not conclusive and other processes such as enhanced specific heat could result in the same effect.

\item  We found that 
\xte\ cannot be explained with a standard crustal cooling model.
It requires additional heat sources located in the outer crust, at $\bar\rho \sim(1-5) \times 10^{10}$\,g\,cm$^{-3}$, releasing  $q\sim(0.1-0.25)$~MeV\,nuc$^{-1}$. 
In addition, we explored alternative scenarios, such as residual accretion during quiescence. 
Even though this model can explain data, the thermal component outside flares is probably not caused by ongoing low-level accretion \citep{Fridriksson2011}.
We also probed the scenario of suppressed electronic thermal conductivity in a thin layer due
to a buried magnetic field. We found that the layer must be thin,  
($\Delta r\sim (12-20)$~m), and located at $\rho\sim 10^{10}$\,g\,cm$^{-3}$.
For a better description it is necessary to solve a 2D problem considering the magnetic field geometry. 

\item The quiescent emission of IGR\,J17480$-$2446 challenges our current understanding of crustal cooling since  its thermal flux  still remains above the value measured in the previous quiescent phase  
after spending 2.2~years in quiescence. This is difficult to reconcile with its short outburst (which lasted only two months). In agreement with \cite{Degenaar2013}, we found that it is 
possible to explain the data if we consider that \tc\ is higher than the one measured in the last quiescent phase. Another possibility are, again,  additional heat sources, 
but in this case, they must be considerably more intense ($q\sim3$~MeV\,nuc$^{-1}$) and must be located in deeper layers ($\bar \rho \sim 10^{11}$\,g\,cm$^{-3}$) than for \xte.

\end{enumerate}

 \begin{figure}
    \centering
    \resizebox{0.95\columnwidth}{!}{\includegraphics[angle=-90]{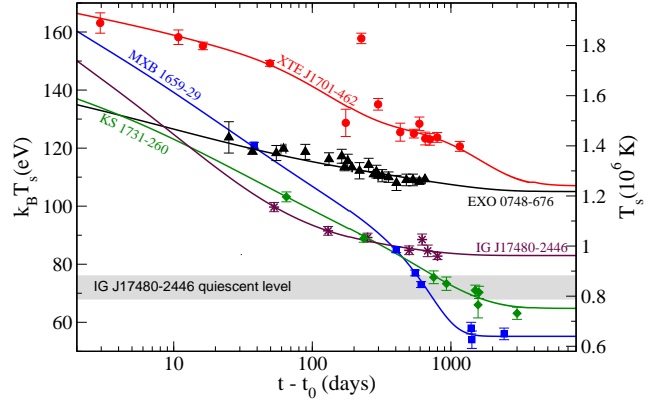}}
       \caption{Cooling curve comparison of all the sources. In each case we show the best fit corresponding to the models:
       crustal cooling (\mxb), crustal cooling with  small energy gap for neutron superfluidity (\ks), crustal cooling (all data fit, \exo),
        crustal cooling with additional heat sources (\xte), and crustal cooling with a high \tc\ (\ig).}
          \label{all}
    \end{figure}

In Fig.~\ref{all} we summarize our results, showing the best fits obtained for all the sources. According to this, 
\mxb\ and \ks\ have already reached thermal equilibrium, with surface temperatures at $kT_{\rm eff} \sim 55$~eV and $kT_{\rm eff} \sim 65$~eV, respectively. \exo\ and \ig\ seem likewise close to equilibrium levels, with 
temperatures of $kT_{\rm eff} \sim 105$~eV and $kT_{\rm eff}\sim 83$~eV, respectively. On the other hand, 
\xte\ is still far from thermal equilibrium, which we predict will be reached in several years at  the value  of $kT_{\rm eff} \sim 107$~eV. 
This high value is similar to that of \exo, while the other sources level off at much lower temperatures.
 \xte\ has an  early observation, which provides valuable information about the
position and intensity of heat sources in the outer crust.
Instead, we do not have information before $\sim$30~days for the other sources. 
An open question is whether the other sources showed an early behavior similar to that of \xte, for which  observational data are lacking.

\begin{acknowledgements}
This research was partially supported by CONICET, 
PIP-2011-00170 (DNA) and by the grant AYA 2010-21097-C03-02 (JAP). 
DNA thanks Physics Dept. of Ohio University where part of this project started 
and Dept. of Applied Physics of Alicante University for warm hospitality. 
\end{acknowledgements}

\appendix
\section {Comparison with previous works}
\label{App_Comparison}
\subsection{\mxb\ results with BC09}

\begin{figure}[!h]
    \centering
    \resizebox{\columnwidth}{!}{\includegraphics[angle=-90]{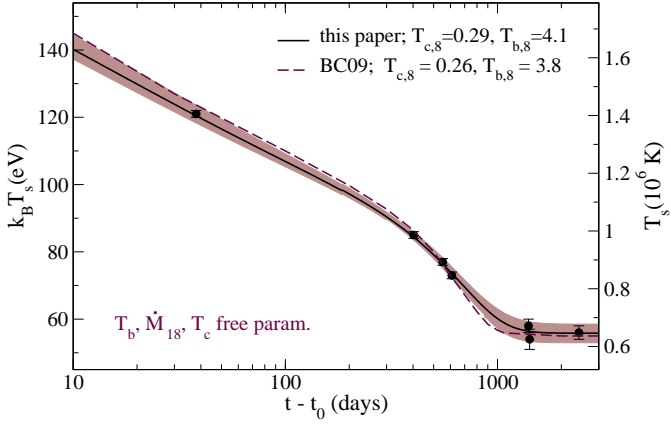}}
       \caption{Comparison of our cooling curves for \mxb\ with BC09.
The solid line is our best fit with $\chi^2=0.54$, 
the dashed line is the result of BC09.       
The brown region is our curves with $\chi^2<2$ corresponding to
$T_{\rm c,8}=0.26-0.32$ and $T_{\rm b,8}=3.9-4.4$. For all curves $\dot M_{18}=0.1$ is fixed.}
          \label{figb1}
    \end{figure}

To check our numerical approach and code, we first compared our results for \mxb\ with BC09, in 
which \dm, \tc,  \tb\ and \q\ are free parameters (Fig.~\ref{figb1}).
Similarly as they did, we fixed $M=1.6$~M$_{\odot}$, 
$\dot{M}_{18}=0.1$ and $Q_{\rm imp}=4.0$  and explored the behavior of the cooling curves against the variation of 
\tc\ and \tb.
In Fig.~\ref{figb1} the solid line corresponds to the best fit obtained with $T_{\rm c,8}=0.29$ 
and $T_{\rm b,8}=4.1$ ($\chi^2=0.54$). The brown zone is $\chi^2<2$ with 
parameters in ranges of $T_{\rm c,8}=(0.26-0.32)$ and $T_{\rm b,8}=(3.9-4.4)$.

We found that observations can be well described by our cooling curves, and 
they  agree very well with BC09 results (dashed line). 
     
\subsection{\ks\ results with BC09 and Sht07}  

\begin{figure}[!h]
\centering
\subfigure[The solid line is our curve with $Q_{\rm imp}=2$. The dashed curve is from 
Sht07.]
{\includegraphics[scale=0.3,angle=-90]{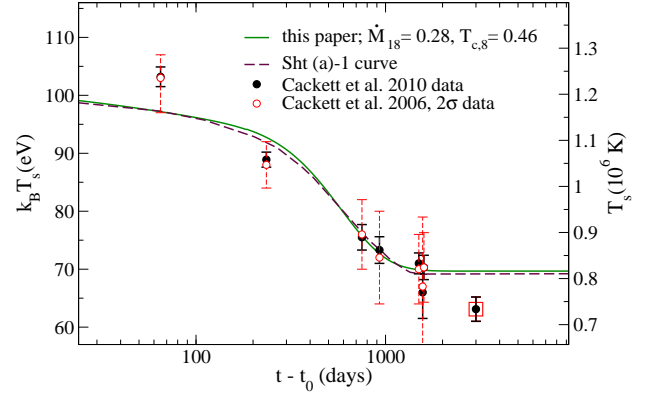}}
\subfigure[The solid curve is our best fit  with $\chi^2=0.5$. Fixed parameters are $\dot M_{18}=0.1$ and $Q_{\rm imp}=1.5$.
The brown region corresponds to $\chi^2<2$ with free parameters varying in ranges
$T_{\rm c,8}=(0.45-0.51)$ and $T_{\rm b,8}=(2.5-3.4)$.]
{\includegraphics[scale=0.3,angle=-90]{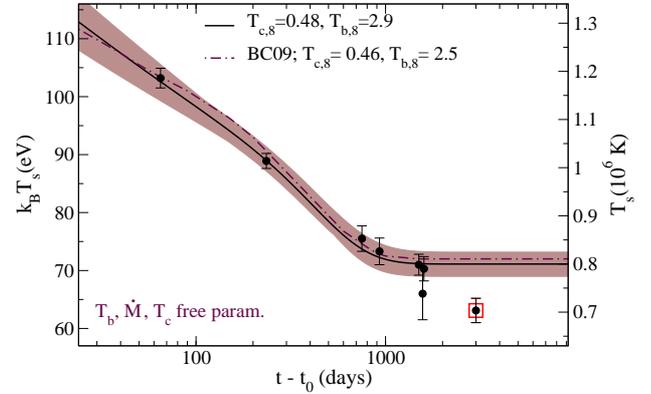}}
\caption{Comparison of our results for \ks\ with Sht07 (upper panel) and BC09 (bottom panel).
Solid curves are our fits and dashed and dotted dashed curves are Sht07 and BC09 results, respectively. 
$M=1.6$~M$_{\odot}$ is fixed for all. The last observation (with a red square) was reported after the publication of those works. }
\label{figa2}
\end{figure}

We compare in Fig.~\ref{figa2} our cooling curves for \ks\ with previous results of Sht07 (top panel) and 
BC09 (bottom panel). 
The dashed line at the top panel is taken from from Sht07, the solid line is our result obtained by letting \tb\  evolve freely (as in Sec.~\ref{reviewMXB}). 
We considered \dm, \tc\ and \q\ as free parameters and found that the data can be explained with the values $\dot M_{18}=0.28$, $T_{\rm c,8}=0.46$, and  $Q_{\rm imp}=2$. 
The NS mass is $1.6$~M$_{\odot}$ and the neutron superfluidity energy gap in the crust is that of \cite{Wambach1993} 
(moderate-superfluidity case in Sht07).
For a better comparison of the results we show in addition to the data of \cite{Cackett2010} (considered so far)
observations from \cite{Cackett2006} (open circles) with 2$\sigma$ error bars, which are the ones considered by Sht07. 

In the bottom panel we compare our results with BC09.  Now  \tb\ is fixed during outburst to a constant value (as in Sec.~\ref{reviewMXB}),  
and we simulated with \tc, \tb\ and \q\ left free to vary.  The  dotted-dashed line was taken from BC09, the solid curve is our best 
fit with $\chi^2=0.5$. The brown region corresponds to $T_{\rm c,8}=(4.5-5.0)$ and $T_{\rm b,8}=(2.5-3.2)$ with $\chi^2<2$; all these curves consider $Q_{\rm imp}=1.5$ (the same value was used in BC09). 
The NS mass was fixed to $1.6$~M$_{\odot}$, and we considered the Sch03 energy gap for neutron superfluidity in the crust.

We conclude that our curves agree well with previous results, which make us confident in our work.
     
\bibliographystyle{aa}	
\bibliography{aa}

\end{document}